# Quantum origin of life: methodological, epistemological and ontological issues

Juan Campos Quemada
Universidad Complutense. Madrid. España

**Abstract**

The aim of this essay is to analyze the role of quantum mechanics as an inherent characteristic of life. During the last ten years the problem of the origin of life has become an innovative research subject approached by many authors. The essay is divided in to three parts: the first deals with the problem of life from a philosophical and biological perspective. The second presents the conceptual and methodological basis of the essay which is founded on the Information Theory and the Quantum Theory. This basis is then used, in the third part, to discuss the different arguments and conjectures of a quantum origin of life. There are many philosophical views on the problem of life, two of which are especially important at the moment: reductive physicalism and biosystemic emergentism. From a scientific perspective, all the theories and experimental evidences put forward by Biology can be summed up in to two main research themes: the RNA world and the 'vesicular theory'. The RNA world, from a physicalist point of view, maintains that replication is the essence of life while the 'vesicular theory', founded on biosystemic grounds, believes the essence of life can be found in cellular metabolism. This essay uses the Information Theory to discard the idea of a spontaneous emergency of life through replication. Understanding the nature and basis of quantum mechanics is fundamental in order to be able to comprehend the advantages of using quantum computation to be able increase the probabilities of existence of auto replicative structures. Different arguments are set forth such as the inherence of quantum mechanics to the origin of life, the quantum nature of matter and the potentiality of quantum computation. Finally, in order to try to resolve the question of auto replication, three scientific propositions are put forward: Q-life, the quantum combinatory library and the role of algorithms in the origin of genetic language.

(Keywords: Quantum mechanics, origin of life, philosophy of science, RNA world, quantum computation, information, emergency, reduction)

**Resumen**

En el presente ensayo se analiza la propuesta sobre la especificidad cuántica del origen de la vida, que se ha convertido en un programa de investigación novedoso sobre el problema del origen de la vida durante la última década. El trabajo se divide en tres partes bien diferenciadas: la primera dedicada al problema de la vida tratado desde el punto de vista de la filosofía y la biología, la segunda en la que se exponen los fundamentos de la teoría de la información y de la teoría cuántica como base conceptual y metodológica para poder abordar, en la tercera parte, los argumentos y las conjeturas sobre un principio cuántico del origen de la vida. De las muchas visiones filosóficas del problema de la vida, existen dos orientaciones filosóficas que influyen poderosamente en las investigaciones actuales: el reduccionismo de corte fisicalista y el biosistemismo emergentista. Desde el punto de vista científico las teorías y evidencias experimentales que aporta la biología quedan resumidas en dos proyectos de investigación sobre el origen de la vida: el mundo del RNA y la teoría compartimentalista. El primero, bajo una posición fisicalista, sostiene que la replicación es la esencia de la vida y el segundo, de inspiración biosistémica, antepone el metabolismo. Los resultados expuestos sobre la teoría de la información desvelan el problema de la improbabilidad de la emergencia espontánea de la replicación. Entender los fundamentos y naturaleza de la mecánica cuántica es esencial para comprender las ventajas que presenta la computación cuántica sobre la clásica a la hora de aumentar la probabilidad de la aparición de la primera estructura autorreplicativa.

Para finalizar, se exponen los argumentos que esgrimen los autores a favor de una especificidad cuántica del origen de la vida, la naturaleza cuántica de la materia y la potencialidad de la computación cuántica, y se concretan tres propuestas científicas que pretenden resolver el enigma de la emergencia de la autorreplicación: la Q-vida, la biblioteca combinatoria cuántica y el papel de los algoritmos cuánticos en el origen de los lenguajes genéticos.

## INTRODUCCIÓN

La siguiente investigación tiene como objetivo analizar la relevancia de la propuesta de especificidad cuántica para el estudio del origen de la vida.

El problema de la emergencia de la vida surge de la determinación de sus características, de los rasgos distintivos de la vida y del estudio de los casos fronterizos. Las diferentes aproximaciones científicas y filosóficas a este problema se abordan desde el biosistemismo, el escepticismo, el funcionalismo, el mecanicismo fisicalista y maquinismo, el esencialismo y el antiesencialismo. Otra de las cuestiones que dificultan la resolución del problema del origen de la vida, es la imposibilidad de encontrar una definición en la que poder alojar a todos los organismos vivientes. Si aceptamos que la unidad de la vida es la célula, entonces se deben estudiar sus componentes y las relaciones y procesos que facilitan la homeostasis celular. La reproducción, replicación y el metabolismo son las funciones esenciales para entender el origen de la vida. El proceso del cuál emergió la vida pudo ser lento y secuencial o surgir instantáneamente de una cualidad emergente del mundo abiótico. Los experimentos que sirven como evidencias de contrastación proceden de la genética, la geología, la paleontología, la astrofísica molecular y, sobre todo, de la síntesis abiótica de aminoácidos y nucleótidos en el laboratorio. Para dar cuenta de los resultados experimentales, las teorías sobre el origen de la vida se dividen en dos grupos: el mundo del RNA y las teorías compartimentalistas. El primero se desarrolló como consecuencia del descubrimiento de la estructura, naturaleza y universalidad del material genético. Propone que la replicación es la clave para entender el origen de la vida. Tras el descubrimiento de la ribozima, un ácido nucleico capaz de llevar a cabo tareas de síntesis de proteínas y replicación genética, se pensó que se había encontrado el replicador primigenio. Pero, aunque la ribozima pudiera autorreplicarse, era tan compleja que se considera altamente improbable la síntesis abiótica natural, es decir, el ensamblaje contingente de los monómeros del caldo primitivo. Además, a este problema se añadía la dificultad de superar la denominada catástrofe del error: relacionada con la necesidad de que los errores en la réplica sean mínimos para que no degenere el protorreplicador en las moléculas que lo forman. Ante tal dificultad, se retomaron los trabajos de Oparín en los que conjeturaba que el origen de la vida se basaba en el metabolismo y la reproducción de vesículas micélicas. Por lo tanto, la vida tuvo que evolucionar desde estas células primitivas a las que conocemos en la actualidad.

De las teorías expuestas hasta la fecha, se desprende que el papel de la replicación y la precisión en la síntesis de proteínas son fundamentales para estudiar el origen de la vida. Esto evidencia que el almacenamiento y procesado de la información es la clave que sostiene cualquier teoría del origen de la vida. Por eso es imprescindible estudiar las aproximaciones clásicas y cuánticas del problema de la información. Los procesadores clásicos utilizan el bit cuyo análogo cuántico es el qubit. La eficiencia y fidelidad de la transmisión o réplica de la información clásica encontraron solución en los teoremas de Shanon. Estos teoremas tienen su contrapartida cuántica. Sin duda, los computadores cuánticos pueden hacer lo mismo que los clásicos, pero gracias al principio de superposición y a algunas propiedades exóticas de su naturaleza cuántica, son capaces de mejorar exponencialmente la capacidad de computación. Esta cualidad es la que explotan los autores de la conjetura cuántica del origen de la vida. Argumentan que, gracias a la potencia de la computación cuántica, se puede resolver el problema de la alta improbabilidad del auto ensamblaje del primer replicador que postula la teoría del mundo del RNA. Claro está que, la no trivialidad del papel cuántico se refiere a la capacidad de mantener estados de superposición entre las moléculas del caldo primordial para que la exploración de ese estado de superposición colapse en una de las configuraciones autorreplicativas posibles. Dos son los problemas fundamentales, uno reside en que todavía no se sabe cómo se pudo mantener el sistema aislado durante el suficiente tiempo para propiciar la exploración del estado de superposición antes de que se produjera el colapso debido al ambiente. El segundo reside en que se desconoce el mecanismo que "elige" el estado de configuración adecuado, aunque se proponen soluciones no confirmadas experimentalmente como el denominado "acto irreversible de amplificación". Finalmente los autores proponen tres soluciones cuánticas para abordar el problema del origen de la vida: la Q-vida, las bibliotecas combinatorias cuánticas y los algoritmos cuánticos en el origen de los lenguajes genéticos. La Q-vida es un protorreplicador de la información capaz de llevar a cabo la exploración de los estados de forma altamente eficiente mediante la propiedad de la superposición. Posteriormente, la Q-vida evolucionó al material genético que conocemos en la actualidad. Las bibliotecas combinatorias explotan la capacidad exponencial del algoritmo de Grover para encontrar, en la biblioteca combinatoria de todas las posibles construcciones moleculares, la estructura del primer replicador. Por último, se postula que un algoritmo cuántico fue el origen del lenguaje genético universal.

Dentro del marco establecido en el estudio del origen de la vida, surgen las siguientes cuestiones que se abordan a lo largo del presente trabajo de investigación:

- ¿En qué consiste la propuesta de una especificidad cuántica en el origen de la vida?

- ¿La propuesta cuántica aumenta las posibilidades de un hipotético mundo del RNA como origen cuántico de la vida?

- ¿Qué posición toma la propuesta cuántica del origen de la vida en el debate entre contingencia *versus* determinismo?

- ¿La conjetura cuántica clarifica el problema de la definición del origen de la vida?

- ¿Defiende la propuesta cuántica un concepto unitario o múltiple del origen de la vida? y ¿de qué tipo?

- ¿Es la vida una propiedad emergente para la hipótesis cuántica? De ser así, ¿Qué tipo de emergencia defiende?

Al final del ensayo, se presenta la bibliografía utilizada, así como todas las referencias que se estiman imprescindibles para abordar el estudio sobre el origen cuántico de la vida.

Debido a la dificultad conceptual y matemática de los temas tratados, se ha creído conveniente la presencia de una apéndice compuesto por un glosario y varias secciones dedicadas a: los principios de la mecánica cuántica; a los efectos exóticos de la mecánica cuántica como el efecto túnel, el efecto Zenon y el efecto Casimir; a la deducción de la ley de Stephan Boltzman y el teorema de Shanon; a la composición molecular del medio interestelar y a la teoría del enlace molecular.

# 1. El problema de la vida

## 1.1. Introducción al problema de la vida.

Las siguientes cuestiones plantean de forma sucinta lo que la filosofía y la ciencia denominan "el problema de la vida".

- ¿Qué es la vida?
- ¿Qué procesos físicos, químicos y biológicos están relacionados con el origen de la vida?
- ¿Cuáles son las condiciones para que la vida emerja y persista en cualquier lugar?
- ¿Existe o existió realmente la vida en algún lugar fuera de la Tierra?
- ¿Fue la vida consecuencia necesaria de la evolución del Universo?

Una de las restricciones más importantes para afrontar la investigación sobre el origen de la vida surge de la determinación de sus características, para lo cual los científicos se plantean la necesidad o futilidad de definir con precisión la naturaleza de la vida.

Las propiedades de los sistemas vivos se pueden resumir en seis[1]:

1. La vida muestra orden y estructura.
2. Toma nutrientes y excreta deshechos.
3. Utiliza energía para crecer y desarrollarse.
4. Lleva a cabo reacciones bioquímicas específicas.
5. Responde a su entorno.
6. Se reproduce y se adapta a las condiciones del ambiente.

Como se verá más adelante, todas estas son propiedades que pueden o no presentarse en algunos sistemas que se califican como vivos, por lo que, para cada caso, existen excepciones tanto positivas, es decir que cumplen la propiedad pero no es un sistema vivo, como negativas, no cumple la propiedad pero se le reconoce como vivo.

En lo que respecta a los rasgos distintivos de la vida, sus condiciones necesarias y suficientes son:

1. El metabolismo que trata el procesamiento de la energía y los nutrientes: capaz de distinguir vivo, durmiente, muerto e imposibilidad de estar o haber estado vivo.
2. La *homeostasis* o equilibrio interno en un ambiente variable.
3. La capacidad de portar información para posibilitar la reproducción: información, mutación, presión selectiva del entorno.

Los casos fronterizos más importantes son

1. Los autorreplicadores metabólicamente dependientes (virus y priones).
2. Los superorganismos.
3. La vida artificial débil.

Los rompecabezas de la vida más importantes son:

1. El origen de la vida.

---

[1] JAKOSKY, Bruce. *Science, society, and the search for life in the Universe.* Tucson: University of Arizona Press, 2006.

2. La continuidad o discontinuidad desde lo no vivo.
3. La emergencia.
4. La jerarquización.
5. La vida artificial fuerte.

Para intentar resolver este rompecabezas es necesario un análisis conceptual, epistemológico y metodológico que oriente y aclare la forma en que la biología afronta las soluciones a varios problemas: el origen, perseverancia y evolución de la vida en el Universo, el papel representado por la química, la física y la tecnología y en qué manera responde a un programa reduccionista y teleológico. Cuatro son los conceptos principales relacionados: la reducción, la emergencia, la complejidad y la direccionalidad.

A continuación, se planteará el problema de la vida y se analizarán las influencias que las distintas posiciones filosóficas ejercen sobre la biología. Posteriormente, se tratarán, sin entrar en detalles, algunas cuestiones ontológicas y epistemológicas derivadas de la interdisciplinaridad como el papel que desempeñan en el reduccionismo de corte fisicalista.

**1.2. El papel de la filosofía de la biología en la orientación de la disciplina.**

**1.2.1. El problema de la vida: un diálogo abierto con la historia.**

La biología representa un ejemplo paradigmático de cómo la ciencia actual está en continuo diálogo con la tradición científica clásica. El origen de la vida desde la materia inerte fue tratado por el atomismo[2] griego apoyado firmemente en presupuestos ontológicos que, andando los años, siguen siendo válidos para el fisicalismo y mecanicismo actual. Por otro lado, la ontología aristotélica sostiene, entre otras, las posiciones holistas y finalistas de algunos de los científicos y filósofos de la biología de nuestros días. Aunque como dice el profesor González Recio: "La Vida no podía ser objeto de explicación puesto que cualquier explicación tenía en ella sus raíces; la Vida no fue inicialmente una región de la Naturaleza sino una manera de pensar la Naturaleza"[3]

Asimismo, el profesor González sintetiza las diferencias esenciales en los rasgos sobre la concepción del problema de la vida entre atomistas y peripatéticos[4]:

Atomistas:

1. Relevancia de la causalidad externa.
2. Carácter accidental de la forma biológica.
3. Ausencia de la idea de organismo como identidad irreductible.
4. Explicación de la macroestructura a partir de la microestructura.
5. Las plantas y los animales son efectos mecánicos surgidos en la historia accidental de la naturaleza, desprovistos de leyes morfogenéticas.

Peripatéticos:

1. Relevancia de la causalidad interna.

---

[2] GONZÁLEZ RECIO, José Luis. "Aire, calor y sangre o la vida inventada desde el Mediterráneo". En: GONZÁLEZ RECIO, José Luis (Ed.). *Átomos, almas y estrellas: estudios sobre la ciencia griega.* Villaviciosa de Odón, Madrid: Plaza y Valdes, 2007, pp. 147-200.

[3] GONZÁLEZ RECIO, José Luis. "Aire, calor y sangre o la vida inventada desde el Mediterráneo". En: GONZÁLEZ RECIO, José Luis (Ed.). *Átomos, almas y estrellas: estudios sobre la ciencia griega.* Villaviciosa de Odón, Madrid: Plaza y Valdes, 2007, p.147.

[4] Ibidem, pp.175-177.

2. Carácter esencial de la forma biológica.
3. El organismo como unidad irreductible.
4. Explicación de la microestructura a partir de la macro estructura.
5. Las plantas y los animales son productos teleológicos.

Estas continúan siendo las cuestiones filosóficas principales que dificultan el consenso de las distintas aproximaciones científicas al problema de la vida en cuanto a la naturaleza de los seres vivos.

### 1.2.2. Posturas filosóficas y científicas ante el origen de la vida:

Es conveniente resumir sucintamente las distintas propuestas filosóficas sobre el origen de la vida para poder enmarcar los problemas que surgen al considerar una propuesta cuántica para el origen de la vida. Si dejamos de lado las corrientes vitalistas, la biología y la filosofía abordan el problema de la vida desde distintas posiciones ontológicas, metodológicas y epistemológicas: biosistemismo, escepticismo, funcionalismo, mecanicismo fisicalista y maquinismo, esencialismo, antiesencialismo, etc.

Para el esencialismo la vida no es un tipo de sustancia química fija, las características de la vida como el metabolismo, la reproducción, etc. trabajando al unísono explican las potencialidades causales evidenciadas mediante los experimentos y, por lo tanto, es en estas potencialidades donde reside la esencia de los tipos naturales.

Los antiesencialistas defienden que las propiedades de los sistemas vivos son azarosas, procedentes de una colección abstracta de todas las características posibles, pero a la vez necesarias y suficientes. Por eso desestiman los casos fronterizos. Algunos antiesencialistas argumentan que el abandono de las viejas taxonomías, y por lo tanto el nacimiento de la concepción esencialista, es consecuencia del debate abierto hace doscientos años sobre la posibilidad de crear vida de lo no vivo. Por lo tanto, la propuesta taxonomista es considerar que *la vida tal como la conocemos es la que es y puede ser*[5] y eso significa que tanto la vida en el laboratorio como la extraterrestre quedarían fuera de esta categoría.

Otras propuestas sobre la naturaleza de la vida se muestran escépticas y afirman que la biología puede continuar sin preocuparse por una definición de vida y creen que ésta no puede categorizarse a través de condiciones necesarias y suficientes. Califican como meros "aires de familia", la forma de interaccionar con el entorno, es decir el metabolismo, la autoorganización, el almacenaje y la utilización de la información para la reproducción o la pertenencia a un grupo con la capacidad de evolución.

Algunos se aproximan al problema desde una orientación semántica buscando una definición que especifique el significado de los términos. Por ejemplo, la definición de vida que propone la NASA: "La vida es un sistema químico automantenido capaz de experimentar evolución darwiniana"

Esta definición de vida muestra problemas de índole científico como por ejemplo que la evolución darwiniana sólo se aplica a poblaciones, descartaría otro tipo de reproducciones y operativamente requeriría de millones de años. Tal vez, una muestra de la dictadura mecanicista de los ácidos nucleicos. Además, muestra vaguedad semántica al utilizar palabras como "capaz".

Dichos autores concluyen que encontrar una definición es cuestión de convención lingüística porque los términos están categorizados por el ser humano pero sólo la naturaleza determina su pertenencia. Por eso afirman que hasta que no se encuentre una buena teoría de la naturaleza de los sistemas vivos seguirán existiendo ambigüedades procedentes de especificar un conjunto de propiedades que informan sobre el parecer y no sobre el ser. En palabras de Cleland y Chyba: "[…] insights gained from philosophical investigations into language and logic strongly suggest that the seemingly interminable nature of the controversy over life's definition is inescapable as long as we lack a general theory of the nature of living systems and their emergence from the physical world."[6]

---

[5] JAKOSKY, Bruce. *Science, society, and the search for life in the Universe.* Tucson: University of Arizona Press, 2006, p. 462.

El ejemplo paradigmático que utilizan en sus argumentos es el caso del agua: hasta que la ciencia no encontró una teoría molecular de la materia que empíricamente comprobara la naturaleza del agua, la definición de agua representaba una convención lingüística que describía el conjunto de propiedades que la caracterizaban. Por lo tanto, para estos autores, todavía nos encontramos en el momento de generar conjeturas sobre la naturaleza de los sistemas vivos y someterlas a juicio empírico.

Para el organicismo o biosistemismo *'vida' es la extensión del predicado 'está vivo' […] es una colección, por lo tanto, un objeto conceptual*[7]. De esto se deduce que "estar vivo" es una propiedad emergente de algunos sistemas complejos. Emergencia entendida desde el punto de vista ontológico como aquella propiedad del sistema de la que carecen todos sus componentes. Por lo tanto, reconoce la vida como un nivel emergente surgido de lo químico[8]. En este sentido la *célula* es el biosistema elemental, aunque no todas las células son biosistemas elementales. Es decir, la célula es el sistema mínimo donde emerge el $βιος$ en un instante determinado y esta emergencia es un cambio cualitativo respecto a cualquier sistema prebiótico. Por lo tanto descartan la vida artificial en sentido fuerte (VAF) ya que las propiedades emergentes del $βιος$ son inseparables de las cosas en sí por lo que la VAF crea un claro problema ontológico. Además, los sistemas artificiales sólo estarían vivos por definición lo que supone un problema epistemológico.

En lo que respecta a los casos fronterizos: los virus, los priones, el RNA, las mitocondrias, etc. pertenecen a sistemas moleculares que no son biosistemas. Esto es así debido a que propiedades como la replicación, el metabolismo, etc. no son propiedades necesarias y suficientes para calificar al sistema molecular como biosistema.

En la propuesta materialista biosistémica emergentista, los organismos en su ambiente con sus subsistemas (moléculas, células y órganos) y supra sistemas (población, ecosistema, etc.) son las unidades de la biología[9]. Por lo tanto, se oponen frontalmente a las propuestas que postulan el origen de la vida[10] de forma reduccionista, ya sea en laboratorio o por azar, mediante autoorganización prebiótica. También contradice a quienes piensan que la historia[11] de los biosistemas es imprescindible para entender el problema de la vida pues no añade nada importante sobre la propiedad de "estar vivo" y, por lo tanto, no se puede sacar conclusión alguna de las investigaciones sobre vida artificial en sentido fuerte. En estos planteamientos, los biosistemistas coinciden con la postura epistémica de los taxonomistas para los que la vida como conjunto de las cosas vivas es la vida tal como la conocemos en la Tierra. No obstante admiten un programa de investigación reduccionista débil: la reducción como método para analizar los distintos componentes y propiedades químico-físicas del biosistema, que caracterizan los sistemas moleculares de los cuales emergerán los biosistemas. Aunque, como se verá en la propuesta cuántica del origen de la vida, la mayoría de los científicos confunde la unificación o integración con la reducción. Sobre todo porque no tienen en cuenta que no todas las teorías son unificables[12] - pues deben compartir conceptos y referentes -funciones, variables, hipótesis, métodos, preguntas,…- y fórmulas puente que aglutinen y cohesionen las distintas ciencias.

---

[6] "[…] los elementos que han surgido de las investigaciones filosóficas sobre lenguaje y lógica sugieren de manera clara que la controversia interminable sobre la definición de vida seguirá latente mientras no tengamos una teoría general sobre la naturaleza de los sistemas vivos y su emergencia del mundo físico". CLELAND, Carol y CHYBA, Christopher. "Defining life". *Origins of life and evolution of Biospheres*, 32 (4), 2002, pp. 387-393.

[7] MAHNER, Martín y BUNGE, Mario. *Fundamentos de biofilosofía*. Méjico: Siglo XXI, 2000, p.168.

[8] Ibidem, p.166.

[9] MAHNER, Martín y BUNGE, Mario. *Fundamentos de biofilosofía*. Méjico: Siglo XXI, 2000, p.173-175.

[10] Ibidem, p.169.

[11] Recuérdese que la capacidad de evolucionar no es propiedad necesaria ni suficiente.

La escuela filosófica enfrentada al biosistemismo emergentista es la reducción fuerte propuesta por el fisicalismo - también en su versión maquinista - y el programa funcionalista.

Para los fisicalistas, los organismos son sistemas fisicoquímicos complejos pero regidos por las leyes físicas y químicas. Por lo tanto sus propiedades no requieren de argumentos fuera de la física y la química. El maquinismo afirma la naturaleza mecánica de los organismos. En la misma línea se postula la propuesta reduccionista que afirma la universalidad de la bioquímica en la cual son suficientes las leyes de la física para dar cuenta de todos los fenómenos moleculares. De ello se deduce que la biología molecular y la bioquímica son dos nombres para una misma ciencia: la física de las moléculas. Es decir, sería suficiente determinar los límites termodinámicos, energéticos, materiales y geográficos para establecer las posibilidades de la vida.

Como ejemplo reciente de la posición reduccionista fuerte son clarificadoras las palabras del físico Esteen Rasmussen ante la primera "célula sintética" conseguida en el laboratorio de Craig Venter con el que discrepa en cuanto a lo que se entiende por vida: "Bottom-up researchers, such as myself, aim to assemble life — including the hardware and the program — as simply as possible, even if the result is different from what we think of as life. All of these deeply entrenched metaphysical views are cast into doubt by the demonstration that life can be created from non-living parts, albeit those harvested from a cell".[13]

La posición funcionalista comparte el objetivo de conseguir 'vida artificial' pero su programa es de *arriba-abajo*: parte de biosistemas como la célula para describir las propiedades de lo viviente mediante, por ejemplo, el cambio del programa genético que opera el *hardware* de la célula. Es decir, pretende sintetizar las partes del sistema viviente y luego organizarlas para que desarrollen el comportamiento dinámico de la vida. De esta forma, su hipótesis es que todas las formas de vida son similares a la de la Tierra y descarta otros metabolismos y sistemas de mantener y reproducir la información. Las 'células artificiales' mediante el ensamblaje, la codificación de la información sobre el sistema – y sus errores – y el metabolismo nos recrean los sistemas autocontenidos que intercambian energía con el exterior y evolucionan. El propio Craig Venter reduce todo a lo que denomina 'genoma mínimo': "minimal genome sufficient to support life"[14]

Los planteamientos funcionalistas explican algunas propiedades pero fallan en otras. Por ejemplo, los basados en el metabolismo son capaces de explicar problemas como la vida en estado 'durmiente' pero descartan como vivientes otros sistemas que presentan propiedades metabólicas (la llama, las células convectivas y la transformación del hierro en óxido de hierro). Lo mismo sucede con los funcionalistas que definen la vida en términos de la autoorganización y automantenimiento – *autopoyesis* –, aunque sabemos que se pueden encontrar sistemas no vivos con estas cualidades (por ejemplo, cristales).

---

[12] MAHNER, Martín y BUNGE, Mario, op. cit. p.142.
[13] "Investigadores de tipo "abajo-arriba" como yo, tenemos como objetivo el ensamblaje de la vida de la forma más simple posible, incluyendo el hardware y el programa, aun cuando el resultado difiera de lo que se entiende por vida. Todos estos puntos de vista metafísicos tan consolidados se tambalean con la demostración de que la vida puede ser creada por elementos no vivos aunque recogidos de una célula". RASMUSSEN, Steen. "Bottom-Up will be more telling" *Nature*, 456, 2010, p. 422. [http://www.nature.com/nature/journal/v465/n7297/pdf/465422a.pdf. Consultado por última vez 04/09/2010].
[14] "Genoma mínimo suficiente para mantener vida". BEDAU, Mark A. "What is life?" En: SARKAR, Sahotra y PLUTYNSKI, Anya. *A Companion to the Phylosophy of Biology*. Oxford: Blackwell Publishing, 2000, p.463.

En el siguiente punto se presentan las diversas teorías sobre el origen de la vida con el propósito de enmarcar con precisión la propuesta de un origen cuántico de la vida. Las teorías alternativas, que buscan el primer replicador, evidenciarán lagunas importantes y visiones antitéticas para sostener la especificidad de la propuesta reduccionista cuántica.

**1.3. El contexto científico.**

### 1.3.1. Las células unidades de la vida.

A mediados del siglo XIX, J.M. Scheiden y T. Scham propusieron la teoría celular: la célula es la unidad estructural y funcional de todos los seres vivos. Así pues, las células son las unidades más pequeñas de las que está formada la vida sobre la Tierra. Las características de la célula se pueden sintetizar en las siguientes propiedades[15]:

1. Todas las células provienen de otras células.
2. Toda célula cuenta con su genoma formado por material genético (*DNA, RNA*) que determina la estructura y función de la célula. El genoma es capaz de replicación.
3. El flujo de información en la célula se produce siempre desde el material genético hacia la proteína. Éste es el dogma central de la biología molecular.
4. La célula es un sistema semi-abierto en equilibrio de flujo que utiliza la membrana celular como frontera.
5. La organización dentro de la frontera es mayor que en el exterior.
6. El recurso energético que utiliza la célula se encuentra en forma de ATP (adenosintrifosfato)

Consecuencia de estas propiedades es la capacidad de las células para responder a los estímulos exteriores, para el movimiento, para el intercambio de productos a través de su membrana y para el crecimiento.

Existen dos tipos de célula:

- Las células sin núcleo (células procariotas) con un tamaño medio entre $0,1$ y $1\ \mu m$. A este grupo pertenecen las bacterias (arquea, protea). Son siempre unicelulares. Poseen membrana plasmática, ribosomas y carecen de citoesqueleto. Su material genético se encuentra en el citoplasma que es pequeño comparado con el de las células eucariotas.
- Células con núcleo (células eucariotas) con un tamaño medio entre $10$ y $50\ \mu m$. El DNA de las eucariotas se encuentra en el núcleo, posee intriones y su tamaño es grande comparado con el de las procariotas. Además de la membrana plasmática y los ribosomas, en ellas se encuentran los orgánulos citoplasmáticos (separación del espacio interior) y el citoesqueleto, que le aporta consistencia y le permite la contracción. El dominio de las eucaria lo forman los reinos de animalia, vegetalia, fungi y protista.

### 1.3.2. Reproducción y replicación.

La información[16] sobre la estructura y las funciones de la célula se encuentra archivada en el material genético (DNA y RNA). La molécula de DNA es un polímero con estructura de doble hélice. Está compuesto por desoxirribonucleótidos, es decir, por ribosa, con una estructura cíclica formada por cinco átomos de carbono, ensamblada mediante grupos fosfato y cuatro posibles bases (adenina, guanina, citosina y timina), situadas en disposición lineal entre el armazón fosfoglúcido. La proporción de adenina es igual a la de timina y la de guanina es igual a la de citosina. La secuencia de tres desoxirribonucleótidos, por ejemplo timina, timina, adenina {TTA}, forma los tripletes que se encargan de codificar los aminoácidos, por ejemplo la secuencia {TTA; TTG} da lugar a la leucina que es un aminoácido proteico, y por lo tanto a las secuencias determinadas que dotarán de funciones específicas a las proteínas.

---

[15] PLATTNER, Helmut y HENTSCHEL, Joachim. *Manual de biología celular*. Traducción de Luis Serra, revisión y prólogo Mercé Dufort. Barcelona: Omega, 2001.

[16] En las bacterias $10^7\ bytes$ y en las eucariotas $10^9, 10^{10}\ bytes$.

Las moléculas de RNA están formadas por ribonucleótidos de adenina, guanina, citosina y uracilo unidos por fosfodiéster. El RNA en las eucariotas se muestra con una única cadena por lo que es menos estable que el DNA. La función del RNA es almacenar y distribuir la información genética, extrayendo la información codificada en el DNA para guiar la síntesis o producción de proteínas a los ribosomas[17]. Hay cuatro tipos de RNA:

a. Mensajero (RNAm 2-5%, $10^5 \, a \, 10^6 u$ ): transcripción o copia de la información del DNA y transporte hasta los ribosomas.
b. Ribosómico (RNAr, 80%, $5 \times 10^5 \, a \, 1,7 \times 10^6 u$ ): estructural.
c. Transferencia (RNAt, $5 \times 10^4 \, u$ ): transporta aminoácidos hacia los ribosomas para formar proteinas.
d. Necleolar (RNAn).

Además de las características anteriores algunos RNA tienen funciones catalíticas como las ribozimas[18]

La replicación de genomas se produce a partir de la división celular cuando se separan las dos cadenas del DNA sirviendo cada una de molde para la síntesis de una cadena complementaria. En la replicación encontramos tasas de error realmente bajas ($1 \, errónea \, de \, cada \, 10^9$). Esos fallos representan mutaciones que junto a la selección natural son la base de la evolución.

### 1.3.3. Metabolismo.

El metabolismo es el conjunto de reacciones y procesos bioquímicos que permiten la vida de la célula. Precisa de la participación de *enzimas* que construyan las denominadas rutas metabólicas. Para mantener la estructura y funciones de la célula, ésta utiliza como material energético el ATP que es un tipo de nucleótido que almacena gran cantidad de energía en su enlace entre los átomos externo y central de fósforo. Esta energía no se desperdicia en forma de calor sino que es capaz de producir trabajo, proporcionando energía para los procesos metabólicos.

---

[17] Verdaderas factorías de proteínas.
[18] Los catalizadores son compuestos que aceleran las reacciones químicas. Las enzimas son biocatalizadores cuya forma más común es la de proteína pero algunas reacciones pueden ser catalizadas por los ribozimas que son moléculas de RNA.

## 2. El origen de la vida

### 2.1. Introducción: la chispa de la vida.

Aunque la vida conocida es la vida celular, no todos los autores están de acuerdo en que el origen de la vida coincide con el de la primera célula. Lo cierto es que la mayor parte de los autores se centran en el estudio de los componentes moleculares de las propias células. Parece pues que el problema del origen de la vida se transfiere al problema del origen de una serie de moléculas, con distintos tamaños y funciones, que en combinación explicarían la emergencia de la vida. Más aún, algunos autores buscan la molécula de la vida, aquella portadora de la esencia del misterio que atrapa lo viviente y es capaz de transferirlo a las demás especies moleculares. Tal vez, sea una mirada nostálgica al viejo vitalismo pero en este caso de corte fisicalista.

No obstante, parece claro para la comunidad científica que una sola molécula no puede explicar las propiedades tan complejas que muestra la vida. Por eso, antes de presentar las teorías y los experimentos, es necesario apuntar que la vida surge de propiedades de sistemas y no de propiedades de moléculas maravillosas, aunque sin éstas la vida no tendría lugar.

En este orden de cosas, desde el punto de vista metodológico destaca la diferencia entre las dos posibles aproximaciones al origen de la vida: una orientada desde la propia vida, a través de las ramas del árbol filogenético, y la otra desde la química prebiótica hacia la bioquímica. Pueden representar puntos de vista complementarios pero algunos autores parecen contraponerlos, olvidando que muchas de las especies moleculares no se comportan de la misma forma dentro y fuera de la célula.

### 2.2. Evidencias experimentales.

Las observaciones y experimentos sobre biogénesis se centran en buscar fórmulas y procedimientos para conseguir abióticamente[19] los compuestos moleculares de la vida, entre los que destacan los ácidos nucleicos y las proteínas.

Uno de los experimentos más populares de todos los tiempos es la síntesis de aminoácidos realizada por Stanley L. Miller en 1953 que encontró continuación en los trabajos de Orgel[20] y otros científicos.

Miller preparó una *atmósfera reductora* con metano, amoniaco, hidrógeno molecular y agua. Sometió el compuesto a descargas eléctricas y comprobó que se formaban fácilmente dos aminoácidos sencillos: la glicina y la alanina aunque el rendimiento no alcanzaba el 2%. Aquél experimento parecía mostrar que los ladrillos de la vida se podían obtener en laboratorio.

Pocos años después, Joan Oró sintetizó en una disolución acuosa de cianuro de amonio la base de la adenina[21] con un rendimiento del 0,5%. El bajo rendimiento de esta "sopa fría" se podía mejorar si aumentaba la concentración para lo cual Orgel propuso que se bajara la temperatura hasta congelar la disolución.

Los dos experimentos se realizaron bajo atmósfera reductora pero la evidencia geológica demostró que la atmósfera primigenia era neutra. Una atmósfera reductora hubiera formado una ingente cantidad de neón que debería permanecer en la misma proporción que el nitrógeno tras el paso a la atmósfera oxidante actual. Por otro lado, las rocas más antiguas de la Tierra procedentes de Groenlandia demuestran que hace 3,8 mil millones de años existían rocas compuestas por óxidos de hierro impensables en una atmósfera reductora. Así pues, la atmósfera primitiva era neutra y en estas condiciones era imposible conseguir aminoácidos y bases de nucleótidos.

---

[19] Sin la participación de organismos vivos.
[20] ORGEL, Leslie E. "Prebiotic Chemistry and the Origin of the RNA World". *Critical Reviews in Biochemistry and Molecular Biology,* 39, 2004, pp. 99-123.

[21] Una de las cuatro bases que componen los nucleótidos del DNA.

La solución a este problema procede del medio interestelar y del fondo profundo, oscuro y caliente de los mares. Las fumarolas que se encuentran en los fondos oceánicos son fuentes hidrotermales productoras de sulfuro de hidrógeno, sulfuros metálicos de función tiol ($SH^-$) y azufre ionizado ($S^{-2}$) que, además de aportar una atmósfera reductora, son capaces de catalizar reacciones químicas e incluso precipitar en forma de vesículas gelatinosas, es decir, burbujas naturales que pueden almacenar moléculas. A esto se debe añadir que la hipótesis de un origen termófilo de la vida encuentra argumentos sólidos en la genética, ya que las bacterias que gustan por este tipo de ambiente son mayoritarias en las proximidades de la raíz del árbol filogenético.

Por otro lado, la materia interplanetaria e interestelar se ha mostrado una fuente muy rica en moléculas orgánicas. En el apéndice V se pueden consultar la procedencia de la mayor parte de las moléculas del grupo del carbono que encontramos en el espacio.

Procedentes del medio interplanetario e interestelar, los cometas, asteroides y meteoritos aportan materia a la Tierra en cantidades que oscilan entre cuarenta mil y sesenta mil toneladas de rocas[22] anuales que se acumulan desde hace más de cuatro mil seiscientos millones de años. Estas briznas de polvo y rocas interestelares transportan en su interior moléculas orgánicas, como el cianuro $HCN,$ que se pudieron formar abióticamente en los primeros estadios del sistema solar facilitando o propiciando la aparición de la vida.

Concretamente, la investigadora Zita Martins del Imperial College de Londres encontró uracilo y xantina en los fragmentos del meteorito de Murchingson (Australia 1969). Además, si se tiene en cuenta que el azúcar necesario para formar los nucleótidos puede obtenerse de la síntesis de formaldehido en las nubes moleculares interestelares y que el ion fosfato se encuentra abundantemente en la naturaleza, entonces los componentes químicos de las moléculas de DNA y RNA pudieron originarse en el espacio.

### 2.3. Teorías sobre el origen de la vida.

Las teorías sobre el origen de la vida se centran principalmente en dos propiedades: la autorreplicación y el metabolismo. Los dos paradigmas en liza se contraponen en la prioridad genésica de las propiedades de la vida.

    a. El mundo del RNA prebiótico sostiene que la replicación es la propiedad clave para explicar la vida y, por lo tanto, para dar cuenta de su origen.

    b. El mundo de las "bolsas de basura" o teoría compartimentalista[23] argumenta sólidamente que el metabolismo es previo a cualquier replicación.

#### 2.3.1. El mundo del RNA.

La evidencia experimental de que el aparato genético es universal y por lo tanto debió existir en el árbol filogenético una primera célula de la que procedemos todos los seres vivos del Planeta es el argumento más sólido para considerar la replicación como la propiedad original y fundamental de la vida. Como se verá, esta afirmación no carece de puntos oscuros todavía por aclarar y de grandes dificultades, tanto experimentales como teóricas.

---

[22] SMITH, Caroline, RUSSELL, Sara y BENEDIX, Gretchen. *Meteorites*. London: Natural History Museum, 2009.

[23] LUIGI LUISI, Pier. *La vida emergente: de los orígenes químicos a la biología sintética.* Traducción de Ambrosio García Leal. Barcelona: Tusquets, 2010.

El mismo año en el que Miller realizaba los experimentos para sintetizar aminoácidos, Watson y Crick descubrieron la estructura del DNA. El origen de la vida encontró un nuevo paradigma en el origen de la replicación. Uno de los problemas más importantes para poder diseñar una teoría consistente sobre el origen de la replicación se encontraba en la relación circular que se presenta entre los dos tipos de ladrillos de la vida: las proteínas y la "molécula de la vida", el DNA. Lo cierto es que la réplica de DNA, formado por cadenas de nucleótidos, requería de la participación indispensable de una subclase de proteínas, llamadas enzimas, que son las responsables de aumentar la velocidad de las reacciones químicas del proceso de síntesis del DNA. Pero, como se ha apuntado anteriormente, la síntesis de proteínas se obtiene con la información almacenada en el DNA. Por lo tanto, surgió un problema circular sobre qué moléculas fueron las primeras en formarse en la sopa primordial, si las de las proteínas o las de los ácidos nucleicos.

La ruptura de la fatídica cadena circular surgió gracias a nuevos descubrimientos sobre las funciones de otra molécula indispensable para el proceso de réplica: el RNA. Al igual que el DNA, el ácido ribonucleico está compuesto por nucleótidos, si bien la base timina ha sido reemplazada por el uracilo. Como se ha visto, una de las funciones del RNA es la de transportar la información del DNA a los ribosomas para sintetizar las proteínas. Los estudios sobre el RNA pronto concluyeron que, además de la transcripción o copia de la información contenida en el DNA, el acido ribonucleico podía también desarrollar funciones catalíticas. Es decir, una macromolécula compuesta por nucleótidos podía emprender labores enzimáticas, como las ribozimas descubiertas por Thomas Cech y Sidney Altman, por lo que las enzimas proteicas eran innecesarias. Este descubrimiento revolucionó la teoría que postulaba la existencia de una sopa primitiva de moléculas.

La teoría propuesta por Manfred Eigen[24] había surgido de un experimento en el que el ensamblaje de nucleótidos se conseguía utilizando un catalizador[25] extraído de un bacteriófago. El caldo primordial estaba compuesto por una gran concentración de monómeros encargados de polimerizar un RNA primitivo, o cuasi especie, con capacidad de crear moldes como estructura primitiva de réplica imperfecta. Las cuasiespecies[26] estuvieron sometidas a variación y diversificación que mediante presión selectiva en la pugna por los monómeros dieron lugar a varias especies o familias estables de RNA (ribozimas). Su asociación con enzimas proteicas pusieron en marcha los hiperciclos que contribuyeron a la mejora de los procesos catalíticos y de replicación hasta la consecución de las moléculas especializadas en el almacenamiento de la información DNA.

En palabras de Walter Gilbert: "Se puede considerar un mundo del RNA, donde sólo hay moléculas que catalizan la síntesis de las mismas […] por lo tanto, la primera etapa de la evolución, se desarrolla mediante moléculas de RNA que llevan a cabo actividades catalíticas necesarias para el autoensamblaje a partir de una sopa de nucleótidos"[27].

Ahora bien, la teoría de Eigen adolece de muchos puntos débiles y profundos problemas que aún no han encontrado solución. El estudio de los argumentos en contra de esta teoría es importante para entender la conjetura de los partidarios de un origen cuántico de la vida.

---

[24] EIGEN, Manfred et al. "The origin of genetic information". *Scientific American*, 4, (244), 1981, pp. 88-118.

[25] Polimerasa.

[26] Rimonucleótidos estereorregulares (polimerización no enzimática).

[27] Apud SHAPIRO, Robert. "El origen de la vida". *Investigación y ciencia,* 371, 2007, p.19.

1. Improbable síntesis natural: Se pueden obtener en laboratorio nucleótidos muy sencillos pero hay infinidad de maneras posibles de enlazar los grupos fosfato, el azúcar y las bases contenidas en el caldo primordial. Además, los nucleótidos pueden hidrolizarse con facilidad en la disolución. Por lo tanto, el surgimiento de la molécula de RNA al azar, en estado natural, requeriría de miles de millones de años.
2. La catástrofe del error:
    - El mundo del RNA prebiótico debe generar un aparato de replicación con un grado muy alto de fidelidad, que es muy improbable que emergiera espontáneamente de la contingencia de los encuentros de moléculas abióticas. Es decir, la suma de errores en la réplica pueden destruir con facilidad las estructuras complejas. Cada vez que se copia un número de bits de información ($N$) existe una probabilidad de error ($\varepsilon$) que está acotada superiormente mediante un factor de selección natural ($S$):

        $$N\varepsilon < \log S$$

        "Si la información suministrada es menor que la perdida en cada generación resulta inevitable una degeneración progresiva"[28]. La cota superior para $\varepsilon$ será de $N^{-1}$. Para los organismos modernos se encuentran en ese límite $N = 10^8 \rightarrow \varepsilon = 10^{-8}$. Las pruebas obtenidas en laboratorio apenas superan el valor de $10^{-2}$ por lo que el número de *bits* de información replicados con fidelidad no puede superar los cien. Una cantidad muy baja de información para crear una molécula tan compleja como el RNA. A este error hay que añadirle otras tres posibilidades de fracaso que acotan superior e inferiormente el tamaño de la población. Las posibilidades de error de las dos primeras aumentan con el tamaño de la población y la última aumenta con la disminución del tamaño de la misma.
    - El RNA ineficiente, que nos indica la presencia de tipos de RNA que se replican con más velocidad pero que pierden su función por lo que puede aniquilar la población.
    - El cortocircuito: si al catalizar salta por error a una reacción posterior se arruina la réplica.
    - El *colapso* de la población: alguno de los ingredientes del caldo primordial pueden, estadísticamente, encontrarse en menor cantidad provocando el colapso de los ciclos de construcción de las estructura moleculares.

Si estos problemas no parecen suficientes para desestimar la teoría del mundo del RNA, lo interesante es que aunque se pudieran conseguir las moléculas, se precisaría una secuencia prebiótica precisa para poderlas ensamblar y si esto fuera conocido todavía restaría por averiguar cómo se puede sintetizar una cantidad de copias inmensa, del orden de $10^{13}$ *por litro*.

Después de lo expuesto sobre la improbabilidad del ensamblaje abiótico natural de las moléculas de RNA, puede comprenderse la afirmación de Pier Luigi Lisi: "… en mi opinión, la aceptación del surgimiento espontáneo de una familia molecular tan compleja y sofisticada en su estructura y función no pertenece, al menos en el día de hoy, al ámbito de la ciencia, sino que

---

[28] DYSON, Freeman J. *Los orígenes de la vida*. Traducción de Ana Grandal. Madrid: Cambridge University Press, 1999, p. 41.

equivale a la aceptación de un milagro (lo que muy bien puede predisponer a la aceptación de otros milagros más tradicionales)"[29]

Recientemente han surgido algunas propuestas novedosas procedentes de la modelización computacional[30] que pretenden tender un puente entre el primer RNA contingente y el primer RNA ribosómico.

No obstante, a pesar de presentar unos problemas que parecen insalvables, esta teoría goza de gran aceptación. Tal vez la aproximación al problema del origen de la vida desde las *partículas* elementales a las células es muy atractiva para el colectivo de físicos teóricos que pretenden reducir todas las leyes del universo a una sola ley. Por eso no es de extrañar que los partidarios de un modelo cuántico que explique la vida sean fervientes admiradores del mundo prístino y geométrico del RNA, pues continúan en la línea del antiguo proyecto cuántico comenzado por antecesores tan ilustres como Schrödinguer y von Neumann.

### 2.3.2. La teoría compartimentalista.

La denominada teoría de las bolsas de basura tuvo su origen en los trabajos pioneros de Alexander Oparín[31], en los cuales proponía una secuencia temporal en la que primero se formaron las protocélulas, luego las proteinas y finalmente los genes. Las protocélulas se constituían a modo de pequeños saquitos de micelas[32] en cuyo interior quedaba atrapada una pequeña cantidad de agua[33] cargada de moléculas orgánicas. La acumulación de moléculas orgánicas en el interior de las gotas lipídicas aumentó la probabilidad de que surgieran las proteinas que comenzaron los primeros ciclos metabólicos y terminaron por ensamblar los genes. La cuestión ahora es cómo se produce la transición de las vesículas primitivas a las células modernas. Doron Lancet[34] propone que las protocélulas primitivas se mantuvieron estadísticamente llenas de moléculas de tamaño reducido entre las que se encontraban *catalizadores*. La incorporación de más moléculas a través de la membrana semipermeable, pudo provocar la división de la bolsa en dos, lo que representaba un tipo primitivo de reproducción. La información trasladada de bolsa a bolsa consistía en una lista de elementos a la que se denominó genoma composicional. Se sabe que la reproducción no requiere la exactitud de lo replicado, simplemente un parecido con la célula madre primitiva. Sólo cuando apareció la célula moderna, surgió la relación conjunta entre la reproducción y el genoma. El fenómeno estadístico de la reproducción pudo generar cadenas diferentes de tipos de bolsas de basura que entraron en competencia sometiéndose a la selección natural. Por lo tanto, la propuesta compartimentalista conjetura un inicio de la vida donde las macromoléculas formadas por péptidos catalíticos prebióticos son atrapadas por moléculas membranogénicas. En estas vesículas se produce un metabolismo con automantenimiento que puede dar lugar mediante la acumulación de lípidos a la autorreproducción. La síntesis postrera de enzimas membranogénicas desde el interior vesicular[35] pudo conducir a una vía muerta o a la síntesis de RNA primordial, que procuró, a su vez, la síntesis de DNA y la autorreplicación interna macromolecular de núcleo y membrana.

---

[29] LUIGI LUISI, Pier. *La vida emergente: de los orígenes químicos a la biología sintética.* Traducción de Ambrosio García Leal. Barcelona: Tusquets, 2010, p. 56.

[30] WORKSHOP OQOL'09: Open Questions on the Origins of Life: San Sebastián-Donostia, May 20-23, 2009: Book of Abstracts (First Draft). Pier Luigi Luisi & Kepa Ruiz-Mirazo (org.)

[31] OPARÍN, Alexander. *El origen de la vida*. Versión al español de Domingo Orozco M. Barcelona: Grijalbo, 1974.
[32] El conjunto de moléculas suspendidas en un líquido coloidal. Una micela típica forma en disolución acuosa una cabeza hidrofílica y una cola hidrofóbica. Está compuesta por lípidos.
[33] Coacervado.
[34] WORKSHOP OQOL'09: Open Questions on the Origins of Life: San Sebastián-Donostia, May 20-23, 2009: Book of Abstracts (First Draft). Pier Luigi Luisi & Kepa Ruiz-Mirazo (org.).

[35] Autopoiética.

La existencia de una membrana semipermeable que delimite lo vivo y lo inerte es primordial para establecer un metabolismo de automantenimiento. Crear orden interior requiere disminuir la *entropía*[36] y esto se consigue mediante transferencia de energía[37] con el exterior, lo que genera desorden exterior y, por lo tanto, aumento de entropía fuera de la membrana. Esta trasferencia de energía es pues la fuente activadora de la autoorganización. Como se deduce de lo dicho, los primeros procesos debieron estar controlados por la termodinámica del sistema que dependía fuertemente de las condiciones contingentes iniciales de cada vesícula. Debido a la alta concentración de moléculas en las protocélulas, aparecieron las enzimas, controladoras eficaces de la velocidad de las reacciones, propiciando el cambio de director en el metabolismo. La transición del control metabólico de la termodinámica a la cinética química fue trascendental para la aparición de las macromoléculas de RNA.

Los autores presentan variaciones sobre las conjeturas del modelo compartimental. Así, por ejemplo, Cairn Smith[38] propone que el entramado regular que presentan los silicatos de arcilla puede catalizar reacciones de superficie proporcionando un patrón para el metabolismo de aminoácidos y proteínas. Así pues, primero sería la arcilla, las enzimas vendrían en segundo lugar para fabricar membranas y las células primitivas resultantes utilizaron arcilla para la reproducción y el metabolismo primitivo. La presión selectiva derivada de la competencia entre células reproducidas por arcilla y células que habían logrado sintetizar RNA, gracias al metabolismo metálico, concluyó con la supremacía de las células modernas.

Existen variantes de esta teoría adaptadas al mundo termófilo de las fumarolas hidrotermales del fondo de los océanos. Éste es el caso expuesto anteriormente de los sulfuros metálicos estudiados como catalizadores primitivos conjeturado por el grupo de Wächtershauser.

Otra de las teorías más conocidas es la expuesta por Lynn Margulis que encuentra su antecedente en Konstantin Merezhkovsky. En esencia, estos autores proponen que el parasitismo y la simbiosis son los motores de la evolución. Algunas funciones de la célula proceden de la incorporación de material genético. Proponen que el RNA es el material genético incorporado y propagado en sucesivas generaciones.

Para finalizar, son interesantes las teorías autopoiéticas que también encuentran acomodo en la compartimentación, ya que, para autores como Pier Luigi Lisi, toda la vida es de tipo celular y el sistema semiabierto de las protocélulas proporciona el fundamento de la homeostasis, o equilibrio entre las reacciones interiores y el mundo exterior, necesaria para "capturar el mecanismo que genera la identidad de lo viviente"[39].

---

[36] $S = \frac{\Delta Q}{T}$. En los procesos reversibles la entropía crece.

[37] En forma de luz, proveniente de reacciones redox, generadas por gradientes electrostáticos, etc.

[38] Apud DYSON, Freeman J. *Los orígenes de la vida.* Traducción de Ana Grandal. Madrid: Cambridge University Press, 1999.

[39] LUIGI LUISI, Pier. *La vida emergente: de los orígenes químicos a la biología sintética.* Traducción de Ambrosio García Leal. Barcelona: Tusquets, 2010, p. 219.

## 3. **Biología e Información**

Si las características fundamentales de la vida son el metabolismo y la replicación, entonces es evidente la gran importancia de la información en biología.

Como se puede deducir de lo dicho, el grado de precisión para realizar estas tareas es trascendental, por eso la ligadura entre orden y probabilidad debe explicar la alta organización biológica en contraposición al desorden del mundo abiótico. Ahora bien, el grado de desorden del sistema es la interpretación estadística del concepto de entropía postulado por Boltzman[40]. La teoría de la información de Shannon, que se expondrá en el apartado siguiente, tiene una estrecha relación con el concepto de entropía. De hecho, el proceso lógico para deducir la ley de Boltzman es equivalente a intercambiar la entropía $S$ por la información[41] $I$ y la *probabilidad del sistema* $W$ por la probabilidad matemática $P$ que caracteriza la relación entre los casos favorables y los casos posibles.

La teoría de Shannon es una buena herramienta para estimar la cantidad de información contenida en la estructura de las proteínas o de los ácidos nucléicos, aunque el uso de este concepto de información no agota, ni mucho menos, todas las posibilidades que ofrece la información biológica. De hecho, la literatura científica califica de causal a la información entendida bajo la teoría de Shannon, es decir, establece una relación entre contingencia y correlación para poder hacer predicciones. Bajo este punto de vista, la información no es explicativa sino inferencial[42].

Para calcular la información que porta una proteína se necesita saber la frecuencia que presentan los distintos aminoácidos que la compone y, con este dato, obtener la probabilidad de que un tipo de aminoácido tome una determinada posición en la proteína. De esta forma, se puede medir el contenido de información por monómero y por lo tanto la información total de la proteína será la suma de los monómeros[43]. La misma idea se sigue para calcular la información contenida en el DNA cambiando el tipo de monómero. No obstante, la medida sintáctica almacenada en la proteína o en el DNA es menor que la medida con significado, ya que todas las combinaciones posibles no tienen por qué transmitir un mensaje biológico relevante. Mediante esta técnica, y teniendo en cuenta la dificultad para caracterizar el contenido semántico de la información, se estima que la cantidad de información contenida en el genoma varía entre $10^2$ y $10^{12}$ bits.

Este problema, que surge al evaluar el contenido semántico, conduce a la revisión del concepto de información utilizado por Shannon. Una información que, además de predecir, explique cómo los sistemas biológicos son capaces de realizar los procesos naturales.

Hay dos formas de explorar nuevos conceptos de información:

1. Pensar que las estructuras biológicas son portadoras de información semántica por sí mismas y esta especificidad puede explicar la singularidad de su actuación en los procesos biológicos. Este tipo de orientación nos lleva a explorar conceptos como el de función biológica, esto es, por ejemplo, la función biológica del bronceado es la protección de los rayos del sol y no la de camuflaje.
2. Utilizar modelos de comportamiento semántico para encontrar relaciones de semejanza con los sistemas biológicos e intentar demostrar que esas similitudes aportan información. Este sería el camino que toman los programas de diseño computacional de vida artificial.

---

[40] Ver Apéndice VI.
[41] Usualmente en unidades de bits.
[42] STANFORD Encyclopedia of Philosophy http://plato.stanford.edu [Consultado por última vez 07/09/2010]

[43] GLASER, Roland. *Biofísica.* [Traducción Félix Royo López y Félix Royo Longás]. Zaragoza: Acribia, 2003.

# 4. Teoría de la Información y Mecánica Cuántica

## 4.1. Introducción a la Teoría de la información: el bit.

Para comprender la potencia de cálculo de los sistemas cuánticos es imprescindible conocer la diferencia entre la información clásica y la información cuántica. La información procesada y almacenada en la naturaleza es de tipo cuántico pero se manifiesta experimentalmente de forma clásica.

Toda medida de un sistema físico no es más que una pregunta que se formula a la naturaleza cuya respuesta viene dada por un número real o una colección de ellos. Así, si se pregunta por la temperatura de una habitación, la longitud de una rama o el tiempo de gestación de un mamífero la respuesta será un número con unidades de temperatura, longitud y tiempo respectivamente.

En el caso de que la pregunta pueda contestarse con dos respuestas, por ejemplo una afirmación o una negación, se dice que el sistema de dos valores posee un *bit* de información. Estas dos posibles respuestas o estados de un bit de información se pueden representar de varias formas por ejemplo: Icara> y Icruz>; Iencendido> y Iapagado>; Iverdadero> y Ifalso>; ISi> y INo>. En la literatura físico-matemática se emplean: I+> y I->; I↑> y I↓>; I0> y I1>.

Cualquier pregunta que se realice a la naturaleza se puede traducir en series de preguntas cuyas respuestas pueden ser afirmativas o negativas. Por lo tanto, el número de bits de información que determina el sistema será la suma de los bits de todas esas preguntas. Por ejemplo, en el caso sencillo de conocer los seis resultados del experimento de lanzar seis monedas al azar, el sistema requiere de seis preguntas cuyas respuestas son cara o cruz o lo que es igual: seis bits de información. Si se representa la cara por 1 y la cruz por 0, entonces el sistema se describe mediante una colección de unos y ceros de tipo: I010000>.

En general, el número posibles de estados se relaciona con el número de bits de información, multiplicando por dos tantas veces como números de bits tengamos. Es decir la siguiente ecuación:

$$N_{estados} = 2 \times 2 \times 2 \times 2 \times 2 \dots .2 = 2^{n \text{ bits}} \qquad [1]$$

En el caso del experimento de las monedas $N_{estados} = 2^{6 \text{ bits}}$. También se puede calcular el número de bits en función del número de estados mediante la expresión:

$$n_{bits} = \log_2 N_{estados} \qquad [2]$$

Cualquier número real se puede convertir en binario con cierta aproximación, por lo que si se desea determinar la medida de una variable física no se tiene más que traducir este número al lenguaje binario, por ejemplo: la temperatura de ebullición del agua, 100 ºC, representada en binario es 1100100. Esta colección de ceros y unos se obtiene sin más que dividir por dos el número 100 consecutivamente y tomar los restos – es decir los ceros o unos – de cada división.

Más complicado es determinar el valor de una variable física en una extensión continua de espacio. En este caso, se dividirá el espacio en pequeñas celdas asignando a cada una de ellas un valor numérico que siempre se podrá traducir a binario. De esta forma, representar la variación de temperatura de una plancha de cocina desde el centro a los bordes no requiere más que dividir la plancha en celdas y medir la temperatura de cada una. Si se toma el número de celdas del ajedrez 8x8, habrá 64 números del tipo 1100100. Si se quiere conseguir una mayor precisión, sólo hay que dividir la plancha en celdas más pequeñas lo que aumentará el número de celdas y por tanto la cantidad de colecciones de números binarios.

Para determinar completamente el sistema físico desde el punto de vista clásico es necesario conocer cómo varía éste en el tiempo. Por lo tanto, además de obtener las colecciones de posibles estados clásicos, que como se ha dicho vienen dadas por colecciones de bits, será necesario actualizar en el tiempo esas colecciones. Las leyes del movimiento no son más que las reglas de variación o actualizaciones de las colecciones en el tiempo.

Si se tienen dos sistemas físicos, como por ejemplo dos monedas, se podrá comprobar visualmente si cambian cada vez que se tiran de forma consecutiva: una posible ley de movimiento o evolución sería cara-cara, cara-cara –es decir la estabilidad del sistema–; otra sería cara-cara, cruz-cruz; otra cara-cruz, cruz-cara, etc.

Como se observa, para la descripción de los bits de información y su movimiento no se requiere de la explicación de los medios materiales utilizados en su consecución, basta con facilitar la colección de números correspondiente.

La mayor parte de las leyes fisicoquímicas que utilizan los sistemas biológicos son de tipo clásico, por lo que obedecerán en última instancia a las leyes de la física clásica: concretamente la mecánica y la electrodinámica clásicas. Para describir el estado dinámico de un sistema, además de las leyes físicas de evolución del sistema, sólo se requiere obtener la posición en el espacio $r(x, y, z)$ y la velocidad $v(x, y, z)$.

Es decir, seis números - tres para las coordenadas espaciales y tres para las componentes de la velocidad - y la ley dinámica correspondiente $F(r, p, t)$.

De esta forma se comprueba que los bits clásicos pueden ser representados en el espacio ordinario como *vectores* cuya longitud, dirección y sentido vienen descritas por sus tres componentes.

**4.2.    Del bit clásico al qubit.**

Las fuerzas que rigen los enlaces entre las moléculas son de tipo electromagnético, por eso se utiliza el *electrón*[44], que es el sistema cuántico más simple, como ejemplo para evidenciar las diferencia entre los bit clásicos y los cuánticos o *qubits*.

Como se ha visto anteriormente, los bits clásicos pueden ser representados por vectores. Por ejemplo, el caso de un sistema con dos estados posibles sería I1> y I0>.

La aguja de una brújula es una buena analogía de la representación clásica del electrón. Para realizar el experimento se debe *preparar físicamente el sistema*[45] y luego medir.

Se prepara físicamente el sistema situando una aguja entre los polos de un imán orientado en una dirección cualquiera. Al retirar el imán la aguja queda sometida al campo magnético terrestre que realiza de esta forma la medida. La saeta se moverá hasta alcanzar el equilibrio en la dirección norte-sur. Se dice de esta dirección que es la de mínima energía porque al rotar la energía magnética se transforma en energía de movimiento de la flecha o energía cinética hasta que llega a la estabilidad. Si inicialmente el sistema hubiera sido preparado alineándolo con el polo norte magnético la aguja no se movería, por lo tanto, la transformación en energía cinética o "emisión de energía" sería nula y si hubiera sido preparado en sentido contrario al campo magnético terrestre, emitiría la mayor cantidad de energía posible. Entre los dos límites se puede decir que la ley de emisión de energía sigue un patrón que depende del ángulo subtendido entre la dirección de preparación –imán– y la dirección de la medida –campo magnético terrestre–.

---

[44] Concretamente se utilizará lo que se entiende por el estado de *espín* de un electrón.
[45] Situar el sistema en unas condiciones iniciales conocidas.

Siguiendo con la analogía, se considera el espín de un electrón como una aguja imantada. Si se prepara el electrón con un imán en una dirección y se mide con respecto a otro imán situado en otra dirección, los dos estados de un electrón "clásico" serían I0> y I1> que respectivamente representan la no emisión o la emisión de energía en forma de luz.

Pero el espín del electrón no es una aguja imantada, no es un sistema clásico y la prueba de ello es que la emisión de luz no varía de forma continua dependiendo del ángulo entre los imanes, sino de forma discreta, lo que significa que emite o no emite. Si la orientación del espín es paralela al imán, no detectamos luz y si la orientación es anti paralela el sistema emite luz, pero lo realmente sorprendente es que, en los casos intermedios, esto es, aquellos en que los ángulos de preparación y detección no son anti paralelos o paralelos, la emisión se puede o no producir. Si realizamos el experimento con un electrón varias veces, dependiendo del ángulo entre la preparación y la medición, se obtiene más o menos probabilidad de que se emita luz, pero la cantidad emitida es siempre la misma: un "paquete de energía". Un experimento equivalente al descrito consiste en preparar muchos electrones de forma que no interactúen entre ellos y ver para un ángulo dado cuántos emiten y cuántos no. Se concluye que:

- La probabilidad de que el sistema emita o no luz depende del ángulo entre la dirección de preparación y la dirección de detección o medida, es decir del ángulo que toma el estado intermedio del sistema.
- La emisión de energía no se realiza de forma continua, como si bajáramos la luz de un regulador, sino que se produce de forma discreta en lo que se conoce como *cuanto de luz* o *fotón* y que se determina mediante la fórmula de Plack: [3] $E = nh\nu$ donde *E* es la energía emitida, $\nu$ es la frecuencia de la emisión o "color de la luz", n es un número natural (0,1,2,3,…) que determina la discretización o *cuantización* y $h$ es la denominada constante de Plack $[h = (6{,}6256 \pm 0.0005)10^{-34} J.s]$.

De esta forma se comprueba que el estado de espín no es sólo la flecha apuntando arriba y abajo, es decir un bit de información. Antes de la medida, el espín tiene muchos estados posibles que se pueden representar con infinitas flechas de tamaño unidad que desde el interior de una esfera marcaran con la punta cualquier coordenada sobre la superficie: esta es la caracterización de un bit cuántico o qubit. Como las coordenadas de esa superficie son infinitas, se requerirían un número infinito de bits clásicos para definirlas. Pero al realizar la medida para extraer la información, destruimos los estados cuánticos – qubits – transformándolos en los estados clásicos o bits. Por esta razón todos los sistemas cuánticos que de alguna forma no se pueden mantener aislados del entorno quedan convertidos en clásicos.

### 4.3. Perspectiva física de la naturaleza cuántica de la materia.

Antes de continuar con el análisis de las propiedades de la información cuántica se debe explorar la naturaleza cuántica que se atribuye a la materia. Se sabe que en las teorías físicas anteriores a la mecánica cuántica, los modelos físicos conducían a las relaciones matemáticas. Esta secuencia fue alterada por la teoría cuántica pues se estableció una estructura matemática que encajaba con los experimentos cuya interpretación física se mostraba controvertida.

Como se ha visto, la analogía con los modelos macro físicos es un método de exploración que sirve para plantear intuitivamente modelizaciones micro físicas. Uno de los ejemplos más frecuentes se encuentra en la representación de la molécula como un conjunto de bolas, que representarían los átomos, unidas mediante muelles a modo de enlaces químicos. Pero las analogías y las metáforas tienen siempre sus límites de aplicación y deben ser comprobadas mediante la experimentación para no caer en concepciones erróneas.

Siguiendo esta idea, la analogía del espín del electrón como una aguja imantada, no representa el modelo cuántico del electrón. En realidad los estados cuánticos no son flechas o vectores de tres componentes sino, como se acaba de ver, son todos los vectores que apuntan en todas las direcciones de una esfera imaginaria de radio unidad. Esta colección abstracta de objetos es lo que se conoce como *espacios vectoriales*. Aunque por razones de economía del lenguaje se siga llamando vectores a los estados, hay que tener en cuenta que realmente se está hablando de espacios vectoriales abstractos y sólo cuando se realiza la *medida* se pasa a los vectores del espacio geométrico real. Sin tomar conciencia de lo anteriormente expuesto, no se pueden comprender los métodos de la mecánica cuántica. Por esta razón, se deben aclarar los conceptos cuánticos tales como el *estado del sistema*, el papel que juega la *probabilidad* y el concepto de *medida*, pues son esenciales para interpretar correctamente la propuesta de un origen cuántico de la vida.

### 4.3.1. Principio de superposición lineal y dualidad onda corpúsculo.

Como se sabe por los principios de la mecánica cuántica, si dos vectores son posibles estados del espín del electrón en un campo magnético, también cualquier vector suma de estos, o combinación lineal, son posibles estados del sistema. Concretamente, si $|1>$ y $|0>$ son esos dos estados de emisión y no emisión, entonces también se cumple que $a|1> + b|0>$ es un estado intermedio del sistema como el representado por las infinitas flechas o radios de la esfera.

De forma general se puede decir que cualquier estado puede formarse, o descomponerse, a partir de una suma o superposición de estados perpendiculares[46] entre sí, como en el caso clásico de la velocidad de una partícula en el espacio $v = v_x|x> + v_y|y> + v_z|z>$, o como en el cuántico de los dos vectores que representan los dos estados del espín del electrón:

$$|\psi> = a|1> + b|0> \qquad [4]$$

El éxito de la conjetura del estado formado por superposición de estados desde el punto de vista experimental explica todos los casos en los que aflora el fenómeno de las interferencias.

Imaginemos un estanque en el cual al caer, en dos puntos distintos, dos gotas de agua provocan dos ondas circulares respectivamente. Al cruzarse comprobamos que en una región donde se acaban de juntar dos crestas de onda aumenta la magnitud de la cresta, esto es, interfieren constructivamente. Pero al cruzase una cresta y un valle se retorna a la planitud o interferencia destructiva. Como la luz es una onda electromagnética, si se hace pasar un haz de luz por una doble rendija se obtiene un patrón de interferencia sobre la pantalla, o lo que es lo mismo, un patrón que combina de forma alterna zonas sombreadas (interferencias destructivas) con zonas iluminadas (interferencias constructivas).

Éste es un experimento totalmente clásico pero se convierte en cuántico si se debilita la emisión de luz reduciéndola a un goteo de fotones. Éstos impactan con la pantalla dejando una estela de puntos sobre ella. Si se espera el suficiente tiempo, se encontrará el mismo patrón de luces y sombras. Es como si los fotones supieran uno a uno cómo situarse para formar las interferencias o, dicho de otro modo más sorprendente, el fotón "interfiriere consigo mismo" al pasar por el montaje experimental de la doble rendija. El mismo experimento se puede realizar con electrones encontrándose la misma respuesta, lo que significa que la materia muestra un comportamiento ondulatorio y, por lo tanto, se le puede asignar una frecuencia o una longitud de onda para explicar las interferencias : [5] $\lambda = h/p$ siendo $\lambda$ longitud de onda o distancia entre crestas consecutivas y $p$ su cantidad de movimiento que toma valores distintos si se trata de ondas electromagnéticas [6] ($p = \frac{E}{c}$) siendo c la velocidad de la luz en el vacío o, si se trata de materia, [7] ($p = mv$).

---

[46] Ortogonales. Ver apéndice I.

Éste es uno de los experimentos cruciales de la mecánica cuántica y dice que dependiendo de cómo se le pregunte a la naturaleza, es decir, cuál es el experimento de medición, si se trata de una película fotográfica o de un banco de interferencias, la materia y la luz se pueden comportar indistintamente como partícula o como onda.

Como se ha dicho, el principio de superposición explica esta paradoja cuántica mediante la interpretación de $|\psi\rangle = a|1\rangle + b|0\rangle$ como la función de onda del electrón o del fotón.

Como $|\psi\rangle$ es una combinación lineal de todos los estados del sistema, representa toda la información que se puede tener de éste, entonces, la interacción con el aparato de medida puede dar como resultado una medida corpuscular o un patrón de interferencias. Es decir, el resultado del experimento no sólo depende de la partícula sino también del acoplamiento entre ésta y el aparato de medida. Por lo tanto la función de onda de la partícula $|\psi_{partícula}\rangle$ y la función de onda del aparato de medida $|\psi_{aparato}\rangle$, que tiene toda la información sobre el sistema de observación, explicarían de forma completa el resultado de la medición. Como se verá más adelante, el problema de la medida es una cuestión abierta y, en lo que se refiere a la filosofía de la física, uno de los temas más candentes.

### 4.3.2. Principio de interpretación probabilístico.

Cuando se asigna una probabilidad en física clásica, se piensa en el mecanismo causal que obedecen las leyes de la mecánica clásica, en la que todo está determinado conociendo las condiciones iniciales de todas las partículas. Por lo tanto, se usan las probabilidades como sustituto del conocimiento exacto de esas causas. Pero en mecánica cuántica, a pesar de que también existen leyes que dirigen el proceso, esas leyes son puramente probabilísticas. Por ejemplo, no interviene ninguna causa en la emisión de un átomo excitado en un instante dado. Las leyes de emisión son, como se ha dicho, puramente probabilísticas.

Por ejemplo, en un sistema cuántico como el del espín de un electrón en un campo magnético, no es posible predecir con total seguridad si se encontrará emisión al efectuar la medida. Ahora sólo se tiene acceso a la probabilidad de que el sistema emita o no luz y esta probabilidad depende del ángulo entre la dirección de preparación y la dirección de detección. Por lo tanto, toda la información reside en el estado del sistema y éste presenta un carácter intrínsecamente probabilístico. Esta probabilidad es un tanto especial, pues si se tiene dos formas de encontrar un resultado que son independientes desde el punto de vista causal la probabilidad del resultado debería ser la suma de las probabilidades. Pero en mecánica cuántica, los dos sistemas pueden interferir disminuyendo la probabilidad[47] como se ha visto en el experimento de la doble rendija.

En el ejemplo del espín del electrón, las probabilidades de emisión y no emisión están relacionadas, respectivamente, con los coeficientes a y b de los estados del la función de onda $|\psi\rangle = a|1\rangle + b|0\rangle$, de tal suerte que $|a|^2 + |b|^2 = 1$. Es decir la probabilidad de emisión + la de no emisión es la unidad.

Realmente lo que se deriva de los dos principios estudiados, el principio de superposición y el de probabilidad, es que mantienen una estrecha relación con lo que se entiende en mecánica cuántica por medida.

---

[47] Esto es debido a que pueden existir interferencias destructivas que disminuyan la probabilidad clásica. Recuérdese el experimento de la doble rendija.

### 4.3.3. Medir en el mundo clásico *versus* medir en el mundo cuántico.

Cuando se desea obtener una medida en el mundo clásico se hace interactuar el sistema con un aparato de observación u observador, por ejemplo iluminando un objeto con el fin de conocer su posición. La longitud de onda de la luz visible[48] es del orden de la micra ($1\mu m = 10^{-6}m$), esto explica por qué se pueden observar las células con un microscopio óptico ($10 - 50\mu m$) y sin embargo, es necesario sustituirlo por uno de rayos X ($0,01 - 1nm$) si se desea estudiar la estructura interna del DNA ($\sim 2nm$).

En la observación clásica el choque de la luz contra el objeto no altera el estado de movimiento de éste, es decir, no cambia su posición ($q$) o su cantidad de movimiento ($p$). Gracias a esto, se pueden conocer con alto grado de precisión los valores iniciales de la posición ($q_0$) y la cantidad de movimiento ($p_0$) de los sistemas y, por lo tanto, se pueden utilizar las leyes clásicas del movimiento que son función de estas dos variables $F(q, p, t)$, para predecir la trayectoria y la velocidad de las partículas.

En física cuántica es imposible determinar la localización del sistema en un punto del espacio sin perder parte de la información sobre su cantidad de movimiento y viceversa. Si mediante un microscopio[49] se intenta conocer la posición de un electrón en un entorno de una décima del tamaño atómico ($10^{-11}m$), se necesitaría luz de longitud de onda tan corta que la cantidad de momento cinético correspondiente a esa longitud de onda, $\Delta p \sim h/\lambda$, sería del orden de $10^{-24} kg\, m/s$ que al ser transferida, siguiendo la ecuación [7], al electrón de masa ($10^{-30}g$) mostraría una enorme indeterminación de la velocidad: aproximadamente una centésima de la velocidad de la luz ($10^6\, m/s$).

En las conversaciones mantenidas por Niels Bohr y Werner Heisenberg, entre octubre de 1926 y febrero de 1927[50], se abordaron este tipo de experimentos. Las cuestiones debatidas pretendían iluminar el problema de cómo explicar la trayectoria continua que muestra un electrón en una cámara de niebla[51] y cómo hacerla compatible con el problema de encontrar la trayectoria para los electrones que se mueven en torno al núcleo del átomo[52]. Bajo esta controversia surgiría un problema que sigue siendo uno de los más controvertidos para físicos y filósofos.

El problema de la medida se podría formular de la siguiente manera: ¿De qué forma las propiedades de la mecánica cuántica dependen de las condiciones experimentales?

En la física clásica, al situar dos sistemas similares, por ejemplo dos agujas imantadas, en las mismas condiciones iniciales, esto es, con el mismo ángulo, el comportamiento consecuente será idéntico. Pero si los sistemas son cuánticos, como en el caso del espín del electrón, lo único que se puede hacer es preparar los dos sistemas cuánticos de la forma más precisa para que tengan la misma probabilidad de emitir o no. Así se dice que los dos sistemas están preparados en el mismo estado. Ahora bien, al medir se hace interactuar al sistema con el aparato de observación y el valor de la medida se obtiene a expensas de destruir el estado inicial, quedando el sistema en un nuevo estado. Por eso se debe preparar nuevamente el sistema después de cada medida en el mismo estado inicial. En el ejemplo utilizado, esto significa realizar repetidamente el experimento para encontrar ese estado del sistema mediante el método de prueba y error. De ahí que después de reiterados experimentos se pueda obtener el *valor medio de la medida*[53]. Así pues, a diferencia del caso clásico donde la observación aportaba datos en forma de números reales, en el caso cuántico sólo se tendrá acceso a los valores medios esperados de *los observables* que son aquellas variables que se pueden medir.

---

[48] Distancia entre dos crestas de la onda electromagnética.
[49] BOHR, Niels. "The quantum postulate and the recent development of atomic theory". *Nature*, 121, 1928, p. 580-591.
[50] RIOJA NIETO, Ana María. "Los orígenes del principio de indeterminación". *Theoria: Revista de teoría, historia y fundamentos de la ciencia*, 10, (22), 1995, pp. 117-142.
[51] Es un detector de radiación con forma de válvula cilíndrica que presenta una cara de vidrio y un pistón móvil. Sirve para fotografiar el paso de las cargas justo en el momento que se condensan pequeñas gotas de hidrógeno por efecto del calentamiento y movimiento del pistón.
[52] El modelo atómico no pudo relacionar la frecuencia de rotación en torno al núcleo con frecuencia de emisión de luz (espectro de líneas) cuando los electrones saltan entre las órbitas.

Pero además, por lo general, no se pueden medir todas las variables a la vez, ya que el estado del sistema cambiará tras la medida y ese estado no tiene por qué ser compartido por todos las medidas posibles[54]. Utilizando una analogía culinaria, no se obtiene el mismo plato al cocer y luego freír las patatas que al freír y luego cocer las patatas. Las operaciones de freír y cocer no conmutan. Así pues, sólo se podrán medir simultáneamente aquellas variables u observables que compartan ese estado[55].

De ahí que el Principio de Incertidumbre de Heisenberg sea consecuencia de que la medida de los estados de la posición y del momento no comparte un estado común o compatible. Por lo tanto, la precisión $\Delta q$, con la que se mida $q$, que es consecuencia de la dispersión del valor medio obtenida para la posición, y la precisión, $\Delta p$, con la que se mide $p$, que es debida a la dispersión de resultados con los que se mide el valor medio del momento, se encuentran relacionados: si se mide primero la posición se deja la partícula en un estado que es sobre el que se efectuará la medición del momento. Se puede comprobar experimentalmente que cuando se intenta hacer pequeña la incertidumbre en la posición $\Delta q$ aumenta la incertidumbre en el momento $\Delta p$ (o en su velocidad) y viceversa.

En la formulación tradicional de Heisenberg[56]:

$$\Delta p \Delta q \geq h/2\pi \qquad [8]$$

Esta relación de incertidumbre tiene una expresión homóloga para las medidas de energía $E$ y tiempo $t$ [57] : si se aumenta la precisión en el conocimiento de la energía de un sistema, entonces se perderá información sobre el tiempo que tarda el proceso físico en tener lugar y viceversa.

$$\Delta E \Delta t \geq h/2\pi \qquad [9]$$

Esta relación se puede utilizar para estimar el tiempo de plegamiento de las proteínas que es el proceso por el que una proteína adquiere su estructura tridimensional que la capacitará para realizar su función biológica.

---

[53] El valor medio $<F> = P_1 F_1 + P_2 F_2 + \cdots P_n F_n = \sum P_n F_n$ donde $F_n$ es el valor de cada medida y $P_n$ representa su correspondiente probabilidad.
[54] Salvo en un factor de proporcionalidad.
[55] Técnicamente se dice que los operadores que representan a los observables conmutan cuando las variables son compatibles.
[56] La deducción de esta ecuación no es trivial. Ver: MESSIAH, Albert. *Mecánica cuántica: Tomo 1*. [Traducción por Carmen de Azcarate y Jaime Tortella]. Madrid: Tecnos, 1983.
[57] $\Delta E \cong v \Delta p$ y $\Delta t \cong \Delta x/v$

Pero realmente, todavía no queda claro cómo se produce la medida. El problema esconde varias cuestiones nada triviales que esperan solución. Por ejemplo cómo las propiedades clásicas surgen de las cuánticas o, dicho de otra forma, si la física clásica es límite de la cuántica[58]. Como se sabe, para Bohr: "a purely symbolic scheme permitting only predictions, on lines of the correspondence principle, as to results obtainable under conditions specified by means of classical concepts"[59]. Por esto, aunque los fenómenos transciendan la descripción de la física clásica, todos los resultados se deben expresar en términos clásicos. Los primeros intentos de resolución del problema pasaron por el enunciado de distintos postulados.

Niels Bohr propuso un principio de la medida en el que afirmaba[60] que si se mide la cantidad $O$ (*como el momento o giro del espín del electrón*) en un sistema $S$ (*como el del estado del electrón*) en el instante t, entonces $O$ muestra una cantidad en $S$ en instante t.[61] Es decir, la medida se postula como una propiedad relacional oculta, porque el valor de la magnitud estaba ahí antes de medirlo.

Werner Heisenberg reclamaba, a la manera de los positivistas, el mismo grado de validación experimental para los conceptos físicos de las teorías que para sus consecuencias experimentales. Así, propuso que no tiene sentido asignar el valor $q$, a la cantidad $O$ para un estado $S$ en el instante t a menos que se haya medido $O$ y muestre el valor $q$ para el estado $S$ en t.[62] Paradójicamente[63], Einstein se opuso a los dos afirmando la objetividad de la medida frente a la dependencia o subjetividad que mostraban las propuestas de Bohr y Heisenberg[64].

La interpretación de lo que sucedía en el acto de la medida tomó un nuevo giro con la presentación matemática rigurosa de la mecánica cuántica que von Neumann desarrolló en su libro *Mathematical Foundation of Quantum Mechanics*. Von Neumann sugirió que, si se conoce el estado de un sistema en un momento dado y las influencias causales en momentos posteriores, se puede determinar su estado según el sistema evoluciona conociendo el estado inicial gracias a que previamente se ha preparado. Esa preparación se sustenta en el Postulado de Proyección de von Neumann: "Si se acaba de realizar una medición que revela un valor dado del observable, entonces, el estado del sistema inmediatamente después de la medición es el que corresponde al sistema con ese valor exacto para la cantidad medida"[65]

Para von Neumann la medida tiene lugar en dos fases temporalmente diferenciadas:

---

[58] El principio de correspondencia de Bohr justificaba que en los contextos en que $h$ era insignificante, la teoría cuántica coincidía con la electrodinámica clásica. Ver: TORRETTI, Roberto. *The Philosophy of Physics.* Cambridge: University Press, 1999, p.313.

[59] "un esquema puramente simbólico permite solo predicciones, en la línea del principio de correspondencia, como resultado obtenible bajo condiciones especificadas mediante el significado de los conceptos clásicos". BOHR, Niels. Apud TORRETTI, Roberto. *The Philosophy of Physics.* Cambridge: University Press, 1999, p.372.

[60] Las cursivas son del autor.

[61] BOHR, Niels. "Quantum Mechanics and Physical Reality". *Nature,* 136, 1935, pp. 1025-1026.

[62] HEISENBERG, Werner. *The physical principles of the Quantum Theory.* Chicago: Dover Editions, 1930.

[63] Einstein había mostrado su rechazo el empirismo de carácter Lockiano al desarrollar la Teoría de la Relatividad.

[64] EINSTEIN, Albert, PODOLSKY, Boris y ROSEN, Nathan. "Can quantum-mechanical description physical of reality be considered complete?" *Physical Review,* 47, 1935, pp. 777-780.

[65] SKLAR, Lawrence. *Filosofía de la física.* Versión española de Rosa Álvarez Ulloa. Madrid: Alianza, 1994. p. 250

- La primera está gobernada por una ecuación lineal determinista[66] donde el sistema cuántico $S$ interactúa con el aparato de medida macroscópico $O$ para medir la cantidad física $C$.[67] Así, el estado del sistema cambia de $S \rightarrow [S + O]$ conservando la cantidad $C$ debido a que la interacción con el equipo de medida sigue una ecuación lineal determinista[68]. Estos estados $[S + O]$ son los que se denominan *estados entrelazados o enredados* porque los dos sistemas $S\ y\ O$ están íntimamente interrelacionados.
- La segunda presenta un proceso completamente indeterminista que se denomina *reducción o colapso* de la función de onda. En esta segunda parte, los estados entrelazados $[S + O]$ saltan a una de sus posibles configuraciones, en la que sólo se puede evaluar la probabilidad de que obtener el valor $C$. En conclusión, no se puede determinar la evolución del sistema mostrándose éste intrínsecamente indeterminista.

En el caso de interpretar la medición como interacción física, se debería revisar la propuesta de von Neumann ya que el detector es macroscópico y por lo tanto estará compuesto por un gran número de estados microscópicos entremezclados compatibles con el macro estado del sistema macroscópico. Si se mide una de las variables microscópicas del sistema, ésta se correlacionará con el macro estado del aparato de medida.

El proceso de la medida referido se puede generalizar al de dos o más sistemas cuánticos que interaccionan o influyen mutuamente. Estos sistemas presentan un estado común de *entrelazamiento o enredo*. Los estados entrelazados comparten el estado del conjunto y pierden su estado individual. Es como si estuvieran interconectados entre sí desde el momento en el que se enredan hasta que se produce el colapso del estado tras la medida[69]. De ahí que al medir una propiedad como el estado de espín de dos partículas entrelazadas, si en una el resultado es $|\uparrow>$, se sabe inmediatamente y sin medir que el estado de la otra es $|\downarrow>$. Por eso tienen propiedades distintas a las de la mera *superposición de estados*, en el que cada sistema conserva su estado individual no influyéndose mutuamente.

Si se mide uno de los sistemas, se conocerán las propiedades del otro gracias al enredo que mantienen. Lo curioso de esta situación es que la propiedad del entrelazamiento persiste independientemente de la distancia de separación que presenten los sistemas por separado.

Dos propiedades del entrelazamiento son de especial relevancia:

---

[66] La ecuación de Schrödinguer. La linealidad permite tratar mediante proporciones la suma de los términos que surgen del principio de superposición. Recuérdese que la dinámica clásica es no lineal.

**[67] Para un desarrollo riguroso ver apéndice I.**

[68] La ecuación de evolución de un sistema cuántico Schrödinguer es lineal lo que permite el tratamiento matemático del fenómeno de las interferencias.

[69] Para el caso de dos electrones, el sistema cuántico es más rico pues contiene algo más que si se tomaran los electrones por separado ya que se puede construir el estado producto o estado combinado:
$|\psi_{elec1}> \times |\psi_{elec2}> = [a\ |1> + b\ |0>] \times [a'\ |1> + b'\ |0>] = [aa'\ |1_{elec1}1_{elec2}>] + ab\ |1_{elec1}0_{elec2}> + ba'\ |0_{elec1}1_{elec2}> + bb'\ |0_{elec1}0_{elec2}>$ que corresponde al estado en que cada uno de los electrones está confortablemente emplazado en su propio estado: $|1_{elec1}1_{elec2}>; |0_{elec1}1_{elec2}>; |1_{elec1}0_{elec2}>; |0_{elec1}0_{elec2}>$. Pero de forma independiente, se pueden tomar los estados entrelazados en el que cada electrón está en el estado compartido con el otro electrón $\{ab'\ |1_{elec1}0_{elec2}> \pm ba'\ |0_{elec1}1_{elec2}>\}$ éstos son los estados de entrelazamiento.

- Muestra una graduación en lo que se refiere a la fortaleza del enredo ante el colapso.
- El estado de entrelazamiento se puede transferir entre sistemas cuánticos.

De la intensidad o fortaleza del enredo ante el colapso, de su capacidad de transferencia y de la interacción con el entorno depende la resistencia del sistema a la *decoherencia* o colapso y, por lo tanto, a que desaparezca esta propiedad cuántica. De lo expuesto se sigue que para mantener las propiedades de enredo se debe aislar lo mejor posible el sistema ya que cualquier interacción podría provocar la desaparición de las superposiciones coherentes convirtiendo el sistema cuántico en uno clásico.

De lo dicho se concluye que la característica cuántica no depende de la escala de magnitudes con las que se trabaja sino más bien de la imposibilidad de mantener aislados los objetos macroscópicos en oposición a la relativa facilidad de aislar los microscópicos.

La propuesta de un origen cuántico de la vida explota esa capacidad que muestran los estados entrelazados o enredados para guardar, repetir, computar y explorar distintas configuraciones de sistemas complejos compuestos por muchas partículas como las que forman el microcosmos. Por lo tanto, una cuestión transcendental para entender la conjetura cuántica del origen de la vida es cómo se puede evitar el colapso de un sistema cuántico para poder utilizar la potencialidad de cálculo y las posibilidades que nos ofrecen esa gran cantidad de estados cuánticos del qubit. Desgraciadamente, tanto las interacciones externas como la propia fragilidad interna del entrelazamiento pueden destruir la coherencia y, por lo tanto, se hace difícil la observación y mantenimiento de los fenómenos de interferencia o enredo.

Algunas de las estrategias para conservar los sistemas cuánticos aislados pasan por:

a. Reducir la temperatura.

b. Aislar el sistema cuántico en trampas magnéticas.

c. Otros explotan el hecho de que la interacción no es suficiente para la consecución de la medida ya que el estado en el que se encuentra el aparato se ve modificado por esa interacción, de lo que se concluye, que si se quiere observar tanto el aparato de medida como el objeto observado deben estar de alguna forma correlacionados de manera estadísticamente reproducible[70]. Si no hay correlación entonces no se producirá la medida. Esta correlación necesita que, de todas las posible variables del estado del aparato de medida, al menos una no cambie independientemente de la interacción, aunque las demás puedan cambiar. Así se puede entender cómo algunas variables colapsan con la medida y otras no.

El problema de la medida sigue abierto. En la interpretación del multiuniverso, se propone una nueva metafísica del mundo, en la que las superposiciones representan alternativas de universos paralelos. Hugh Everett y John Archibald Wheeler postulan que después de la medición, las componentes de la superposición siguen existiendo, por lo que en cada medida se desdobla en múltiples universos. Los proponentes de una especificidad cuántica para el origen de la vida utilizan en distintas versiones la conjetura de los múltiples universos.

---

[70] En términos matemáticos, podemos decir que antes de la interacción la medida de la energía del sistema, depende linealmente de tres componentes: $H = H_{sistema} + H_{parato} + H_{correlación}$. En: BOHM, David. *Quantum Theory*. New York: Dover Publications, 1989, p 584-590.

Por último, se debe señalar que algunas corrientes de investigación cuántica abordan el problema de dualidad, entre el sistema cuántico y los aparatos de medida, postulando una nueva lógica, pues la lógica expresa las bases que vinculan nuestras propuestas sobre el mundo[71].

En lo expuesto en las líneas precedentes, se aprecia claramente cómo las propiedades cuánticas han alterado de forma considerable aspectos profundos de la comprensión del mundo como la causalidad y el determinismo. No obstante la mayor parte de los físicos optan por las interpretaciones más pragmáticas de Bohr o Heisenberg mientras que otros siguen las propuestas de von Neumann y sus nuevas formulaciones.

**4.4. Computación Cuántica.**

El almacenamiento y procesamiento cuántico de la información se apoya en las características básicas de la naturaleza cuántica de la materia que se acaban de exponer. Utilizando reglas cuánticas se intentan explicar los complejos principios emergentes de los sistemas físicos y de los sistemas vivos.

La conjetura de un origen cuántico de la vida se sustenta en la gran capacidad de explorar distintas alternativas que muestran los sistemas cuánticos para conseguir la replicación, así como la seguridad y fidelidad en la réplica para evitar los errores y su propagación.

Las características esenciales de los sistemas cuánticos se pueden resumir en tres[72]:

i. El estado del qubit se encuentra en superposición de varios estados con sus correspondientes valores.
ii. Un estado no se puede leer, esto es medir, sin que esto signifique la elección de alguno de esos valores.
iii. No se pueden *clonar estados*[73] arbitrarios.

Para almacenar y computar es necesario encontrar un sistema físico donde poder codificar la información. Varios son los soportes posibles: en el espín de varios electrones, en las capas de valencia de un átomo, en el núcleo atómico, en las moléculas, etc. Después se requiere procesar la información mediante la traducción o codificación de la información en bits, para concluir con su medida o decodificación posterior vigilada mediante algún método que asegure el éxito de la operación.

Benjamin Schumacher ha resumido y relacionado estas premisas del proceso en la siguiente pregunta[74]: ¿Qué mínima cantidad de sistema físico necesitamos para procesar la información bajo un cierto criterio de éxito?

### 4.4.1. Teoría clásica de la información.

La teoría clásica de la información[75] resuelve la cuestión de cuánta información se puede transmitir en un canal de comunicación con ruido.

---

[71] SKLAR, Lawrence. *Filosofía de la física.* Versión española de Rosa Álvarez Ulloa. Madrid: Alianza, 1994, p. 285.
[72] En lo que se refiere a computación e información cuántica: NIELSEN, Michael y CHUANG, Isaac. *Quantum computation and quantum information*. Cambridge: University Press, 2000.
[73] El teorema de no clonación de estados cuánticos de Wootters, Zurek y Dieks, (1982) prohibe la creación de copias idénticas de un estado cuántico arbitrario conocido.
[74] SCHUMACHER, Benjamin. "Quantum coding". *Physics Review A.*, 51, 1995, pp. 2738-2747.
[75] SHANNON, Claude. *"*A mathematical theory of communication". *Bell Systems Technical Journal*, 27, 1948, pp. 379-423, 623-565 y SHANNON, Claude. "Communication theory of secrecy systems". *Bell Systems Technical Journal*, 28, 1949, pp. 656-715.

La primera tarea consiste en codificar un canal en ausencia de ruido y después en un canal ruidoso, que es aquél en el que las redundancias alteran el mensaje haciéndolo incomprensible.

Se ha visto que la información clásica de cualquier tipo se puede codificar en colecciones de ceros y unos, es decir en un sistema binario o bit clásico. Si se tiene un alfabeto finito como el del código genético correspondiente a las bases: guanina (G), citosina (C), timina (T) y adenina (A), a cada una de las bases se le designará por una colección de números binarios 010000. Las distintas palabras (incluidos los símbolos gramaticales) del código tomarán la forma GAT, ACT, TTA etc. Cada uno de los elementos de la palabra tiene una probabilidad de aparición $(p(G), p(A), ...)$ que viene dada por la relación entre los casos favorables y los casos posibles y que respeta la regla de Laplace donde la suma de todas las probabilidades es 1. De esta forma se generarán cadenas de mensajes en las cuales las probabilidades no varían si el carácter (G, A, T,) está en una palabra o en otra, ni depende de la posición que ocupe en cada palabra. Si el mensaje tiene $n$ bits de información, se puede reducir la dimensión a $nH(A)$ quitando las palabras redundantes y las que no aportan información adicional. Esto se lleva a cabo mediante lo que se denomina la ley de la entropía de Shannon $H(A)$ [76], que tiende a conservar las palabras improbables aunque posibles y a eliminar las más probables pues son las que tienen mayor facilidad para repetirse. Ahora ya se tiene un conjunto de palabras codificadas en bits para transmitir o replicar. La siguiente cuestión se centra en saber de qué forma llega o se replica esa palabra para obtener una probabilidad alta de llegada o de réplica. Para conseguir una alta probabilidad de acierto se utiliza la probabilidad condicionada que informa sobre lo que llega al receptor o se replica suponiendo que procede del alfabeto de origen GTA... Es decir, se deben eliminar las combinaciones de bits $|1000010>$ que no proceden de nuestro alfabeto origen G, T, C y A. También es preciso especificar cuál es la probabilidad de que el mensaje saliendo del emisor llegue al receptor[77] De ahí se consigue conocer la capacidad de bits de salida perfectamente transmitidos.[78]

Una vez se consigue la transmisión y la recepción del mensaje, es necesario corregir los posibles errores que se hayan producido. Para esto se repite cada bit un número de veces suficientemente alto para descartar fallos en la copia o transmisión. Por ejemplo, en la palabra GAT para evitar el fallo GAA se convierte GAT en → GGGGAAAATTTT, cuantas más repeticiones, menor será el número de fallos pero también se ralentizará el proceso de copia. Como el ritmo de transmisión del código no debe sobrepasar la capacidad del canal, para mejorar el ritmo de copia se utilizan los llamados correctores clásicos de errores (CCCE) que son librerías de palabras a modo de gran diccionario que comparten el emisor y el receptor. De esta forma disminuye el número de repeticiones ya que las palabras son cotejadas en el diccionario para ser admitidas y los fallos se pueden corregir asignando por comparación una palabra del diccionario entre las de mayor proximidad a la recibida[79].

Por último se requiere que el mensaje sea copiado o transmitido con la suficiente fidelidad. Esto significa que se debe obtener la probabilidad de que el mensaje descodificado, es decir una vez que se convierten las colecciones de bits en GTA, etc., coincida con el original.

---

[76] $H(A) = \sum p(a_i) log_2 p(a_i)$ con $0 \leq p(a_i) \leq 1$ y $-\infty < log_2 p(a_i) < 0$. Por un lado el log favorece las palabras con probabilidad baja y disminuye la cantidad de palabras con alta probabilidad pues son las que con facilidad podrán ser redundantes.

[77] $\sum P_{y/x}(y|x) p_x(x_i)$. El primer factor es la condicional de que llegue saliendo del emisor. El segundo representa la probabilidad de que el mensaje de salida sea uno específico, por ejemplo GTCA

[78] $C = \sup Indice (x:y) = \sup \sum \sum P_{y/x}(y|x) p_x(x_i) log_2 \frac{P_{y/x}(y|x)}{p_x(x_i)}$

[79] GALINDO TIXAIRE, Alberto. "Quanta e Información". *Revista española de física*, 14 (1), 2000, pp.30-46.

### 4.4.2 Teoría cuántica de la información

La teoría cuántica de la información se basa en los principios clásicos enunciados para la teoría clásica. La analogía se establece sustituyendo el código en bits por los qubits. Se construye un alfabeto cuántico compuesto por estados cuánticos puros distintos $[|G>;|T>;|A>\cdots]$, codificados en qubits I010000>, teniendo en cuenta sus probabilidades correspondientes $[p_{|G>}, p_{|T>}, \ldots]$ donde se cumple que la suma de las probabilidades sea la unidad. El mensaje estará formado por cadenas de palabras (incluidos los símbolos gramaticales) formadas por letras del alfabeto $[mensaje_{i=}|G>|T>|A>|G>|G>|A>|T>|G>\cdots]$, en el que las probabilidades de ocurrencia de cada una de las letras vuelve a ser independiente de la situación de cada letra en la palabra considerada. En el caso clásico la información venía dada por n bits, que son colecciones agrupadas de ceros y unos. En información cuántica su contrapartida son los qubits que forman parte de un espacio vectorial[80] y por lo tanto pueden superponerse linealmente $a|G>+b|T>+c|A>+d|C>\cdots$ formando estados[81] $|GTA>,|GGC>,|TGA>,\ldots$ o incluso entrelazarse o enredarse construyendo estados comunes del tipo [$\frac{1}{\sqrt{2}}|GTA> \pm \frac{1}{\sqrt{2}}|TGA>$], por lo tanto ahora las colecciones de letras incluye las clásicas y muchas más, por eso la dimensión[82] será mayor que 'n'. De nuevo se puede encontrar un método para evitar palabras redundantes o que no aporten información adicional. La solución cuántica análoga a la entropía de Shannon fue encontrada por Schumacher[83] y lleva el nombre de entropía de von Neumann $S(\rho)$. Representa el número medio de qubits de información cuántica esencial por letra del alfabeto. Siguiendo con las líneas maestras del modelo clásico se propone que, teniendo en cuenta la existencia de estados enredados, también se pueda conseguir una alta probabilidad de llegada al receptor o de replica suponiendo que procede del alfabeto de origen y por lo tanto eliminando las combinaciones de qubits que no procedan de este alfabeto. Esto se consigue mediante la información de Alexander Holevo que, teniendo en cuenta el método de preparación del estado, indica la forma de reducir la entropía de los mensajes de salida cuando se conocen los de llegada[84].

En el caso cuántico el papel de la posibilidad, ritmo y fidelidad de la compresión dependen de que los estados del alfabeto $[|G>;|T>;|A>;\ldots$ sean ortogonales[85].

La teoría de corrección de errores cuántico es un campo en candente desarrollo. Los errores pueden suceder en la preparación de los estados iniciales, en el *hardware* utilizado[86] y en la lectura de los datos al final del proceso.[87]

Es evidente que no se puede ampliar el método clásico de la repetición ya que medir o "leer" un qubit es destruirlo y, además, no es posible copiar fielmente un qubit desconocido[88]. Los errores en los qubits no sólo provienen de la inversión de ceros o unos como los clásicos, sino también de otras fuentes como los coeficientes de la superposición. La solución a este problema se encontró[89] mediante procesos de carácter físico transformadores de qubits que corrigen el error sin leer los estados individuales, es decir, que deben discernir el qubit inapropiado e invertirlo[90].

---

[80] Un espacio de Hilbert con densidad de estados: $\rho = \sum_i p_i |Z_i><Z_i|$

[81] Se han tomado dos para simplificar la notación para que sirva como ejemplo aunque no es exacto. Ver nota 34.

[82] Las colecciones de mensajes con n letras vienen dadas ahora por $\rho^{\otimes n}$ de dimensión $|A|^n = 2^{n\log_2 A}$

[83] GALINDO TIXAIRE, Alberto. "Quanta e Información". *Revista española de física*, 14 (1), 2000, pp.30-46.

[84] No obstante, recuérdese que leer el mensaje significa deshacer el entrelazamiento y preparar un estado no es más que asignarle un operador proyección. Esto es, como colocar un campo magnético en una dirección y sentido determinados para condicionar los estados del espín del electrón. Matemáticamente es tomar el operador sobre el estado $\sum p_i \rho_i$.

[85] En el caso vectorial se diría perpendiculares.

[86] Es decir en las puertas lógicas. Ver nota 94.

[87] SALAS PERALTA, Pedro J. y SANZ SÁENZ, Ángel L. "Corrección de errores en ordenadores cuánticos". *Revista Española de Física (REF),* 20 (1), 2006, pp. 20-27.

A modo de síntesis de lo expuesto sobre información cuántica, se puede decir que además de contener a la información clásica, aporta unas cualidades extra debido a la posibilidad de la superposición de estados y al entrelazamiento. Es decir, además de tener unos y ceros en el mensaje[91], por ejemplo en el espín de dos electrones, se pueden utilizar las superposiciones de estados y los entrelazamientos[92] en los que se muestran todas las posibilidades a la vez.

Por lo tanto, la capacidad de trabajar con superposiciones es la característica principal de cualquier proceso cuántico de trasporte, copia o depósito de la información. En especial se emplea en los denominados ordenadores cuánticos[93], que precisan para su construcción lo siguiente: identificar el sistema físico que provea de los qubits, poder convertir los estados al estado inicial (re-inicialización) y capacidad de lectura, encontrar un sistema bien aislado que evite la disolución o colapso del enredo, inventar algoritmos o puertas lógicas cuánticas[94] que, utilizando el principio de superposición, resuelvan los problemas y, por último, que esas puertas lógicas no se "saturen"[95] cuando el número de qubits se hace cada vez mayor.

Dos son las cuestiones más importantes dentro de la propuesta de un origen cuántico de la vida: la primera es conseguir algoritmos que solucionen problemas complejos con celeridad y la otra es mantener el entrelazamiento para no convertir el proceso en una computación de tipo clásico antes de que se llegue a la lectura de los resultados. Algoritmos cuánticos como el de Grover[96] convierten los problemas clásicos difíciles en fáciles. Es fácil apreciar la diferencia de potencia de cálculo mediante el siguiente ejemplo: encontrar un número de teléfono en un listín (i.e.10.000 entradas) requiere de $N/2$ (5000) operaciones con una probabilidad de éxito de $1/2$ y si se utiliza el algoritmo de Grover sólo se requiere de $\sqrt{N}$ (100) operaciones.

Hay distintas técnicas para evitar el colapso y la mayoría dependen del soporte físico del sistema. Los utilizados en computación cuántica son átomos (preparados mediante luz laser), superconductores (utiliza los estados de pares de electrones) y fluidos en los que los estados de las moléculas están preparados mediante resonancia magnética nuclear (RNM). Éste último caso es un ejemplo relevante de cuáles son las condiciones y los procesos requeridos para realizar computación cuántica con moléculas prebióticas.

---

[88] Teorema de imposibilidad de clonación. Ver: NIELSEN, Michael y CHUANG, Isaac. *Quantum computation and quantum information*. Cambridge: University Press, 2000.
[89] STEANE, Andrew Martin. "Error correcting codes in quantum theory". *Physical Review Letters*, 77 (5), 1996, pp. 793-797.
[90] El código cuántico corrector de errores (CQCE) de Shor y Steane, basado en la propiedad cuántica del entrelazamiento, funciona mediante puertas lógicas (ver nota 94) que dan la vuelta y que cambian el signo del entrelazamiento de un conjunto de nueve qubits entrelazados.
[91] $|1_{elec1} 1_{elec2}>; |0_{elec1} 1_{elec2}>; |1_{elec1} 0_{elec2}>; |0_{elec1} 0_{elec2}>$
[92] ab' $|1_{elec1} 0_{elec2}> \pm$ ba' $|0_{elec1} 1_{elec2}>$} con la correspondiente normalización
[93] CIRAC SASTURÁIN, Juan Ignacio. "Quanta y computación". *Revista española de Física,* 14 (1), 2000, pp.48-52.
[94] Los algoritmos son en esencia listas de instrucciones necesarias para resolver un problema. Los diagramas de flujo de preguntas en los manuales son ejemplos de algoritmos. En el caso clásico la información en bits pasa a través de las puertas lógicas universales que son aparatos físicos que pueden realizar cualquier cálculo mediante la transformación de los bits: la puerta NOT convierte $|1> \to |0> \ y \ |0> \to |1>$ y la puerta XOR $|11> \to |0> \ y \ |10> \to |1> \ |01> \to |1> \ y \ |00> \to |0>$. Por supuesto, tienen su análogo cuántico que son capaces de dar la vuelta a los qubits superpuestos CNOT o alterar el signo de la superposición y también pueden formar estados entrelazados provenientes de estados individuales mediante la puerta CXOR.
[95] Esta es la propiedad de escalabilidad.
[96] GROVER, Lov K. "Quantum Mechanics Helps in Searching for a Needle in a Haystack" *Physics Review Letters*, 79 (2), 1997, pp. 325-328.

El fluido clásico está compuesto de un gran número de moléculas donde cada qubit viene representado por un grupo pequeño de ellas. Es decir, el sistema muestra la capacidad de soportar un número elevado de ordenadores cuánticos. Si gracias a un campo magnético se preparan los estados de los distintos núcleos atómicos de carbono e hidrógeno de las moléculas, por ejemplo de cloroformo ($CHCl_3$), entonces con un fotón en radiofrecuencias se puede cambiar el espín del carbono que evolucionará dependiendo del espín del núcleo de hidrógeno realizando la labor de las puertas lógicas. Gracias a este tipo de operaciones se pueden manipular las propiedades surgidas del entrelazamiento entre núcleos y mantener la coherencia cuántica con más fortaleza debido, como se ha apuntado anteriormente, al número elevado de ordenadores cuánticos.

# 5. Una propiedad cuántica no trivial para explicar el origen de la vida.

## 5.1. Argumentos en defensa de un origen cuántico de la vida.

Un grupo de físicos y bioquímicos[97] se plantean si, más allá de la naturaleza mecano-cuántica que presenta la vida en el ámbito molecular, ciertas propiedades cuánticas de la materia, tales como la superposición, el entrelazamiento[98], el efecto túnel, el efecto Casimir, etc., fueron imprescindibles y determinantes para la emergencia de la vida. Es decir, sostienen que la "chispa" que propició el origen de la vida fue un fenómeno puramente cuántico.

Lo que los autores entienden por "trivial" es la importancia de la física cuántica para dar cuenta de la estructura molecular y de la afinidad química que condiciona la función celular. Conjeturan que la mecánica cuántica permitió la emergencia de la vida desde el nivel atómico, sin la intermediación de la química compleja[99], debido a la naturaleza cuántica que presenta la información biológica.

Las macromoléculas que dieron lugar a la vida son tan complejas que un origen de la vida mediante síntesis abiótica de compuestos sencillos ensamblados de manera contingente es imposible. En palabras de Davies: "It will in fact be a near miracle"[100]

Entiende Davies[101] que no hay diferencia entre un milagro y un suceso que es tan improbable que puede haber sucedido sólo una vez en el universo. Por eso cree que la posición científica se sustenta en la existencia de principios físicos que favorezcan la auto organización para convertir la materia en vida.

El principio en el que apoya su análisis es que la vida sigue las leyes de la naturaleza de forma automática por lo que debería existir algún tipo de "principio de la vida", que debería estar contemplado en las leyes de la física. Por lo tanto, la cuestión sobre el origen de la vida se puede dividir en dos preguntas encadenadas:

a. ¿Se conocen las leyes físicas que contemplan el principio de la vida?
b. En caso afirmativo, ¿son estas leyes físicas cuánticas o clásicas?

Los autores argumentan su conjetura en la naturaleza cuántica del mundo y en las propiedades novedosas que surgen de la teoría cuántica como la superposición y el enredo de los estados. Concretamente, sostienen las siguientes afirmaciones:

1. El mundo no es clásico pues la teoría fundamental de la naturaleza es la mecánica cuántica. En palabras de Davies: "if we are looking for a new principle in physics, by default it belongs in quantum mechanics, or else quantum mechanics is not the correct description of the world"[102].
2. No es razonable que la propiedad sea clásica, siendo la vida un producto de la actividad molecular que pertenece al dominio cuántico.
3. Como en mecánica cuántica la observación depende del observador, entonces la conexión entre la vida y la mecánica cuántica se encuentra en el observador. Pero, no porque los estados colapsen[103] debido a la conciencia, sino porque la mecánica cuántica es incompleta debido a su descripción probabilística.

---

[97] ABBOTT, Derek; DAVIES, Paul C. W. y PATI, Arun K. (Eds.). *Quantum aspects of life*. London: Imperial College Press, 2008.
[98] Ver punto 4.3.3.
[99] El modelo del enlace químico, se soporta en la teoría cuántica y encuentra su validación experimental en la explicación de los espectros de emisión y absorción que presentan los distintos elementos.
[100] "De hecho será casi un milagro" ABBOTT, DAVIES Y PATI, op. cit. p. 352.
[101] Ibidem, p. 360.
[102] "Si buscamos un nuevo principio en física, por defecto pertenece al ámbito de la mecánica cuántica, si no, la mecánica cuántica no es la descripción correcta del mundo". ABBOTT, DAVIES Y PATI, op.cit. p. 353.

4. La experiencia ha demostrado que cualquier tipo de tecnología descubierta por el hombre ya se encontraba antes en la naturaleza. Si esto es cierto: "it's hard to believe that nature hasn't (already) made clever use of quantum effects"[104]. Por ejemplo, encontrando subespacios libres de decoherencia[105].
5. Independientemente de la complejidad que presenten los sistemas, pueden existir en superposición. Por lo tanto, al igual que otros sistemas complejos han mostrado estados superpuestos en el laboratorio, entonces no hay razón para que, una vez salvado el problema de la decoherencia, no se puedan superponer los estados de sistemas vivos. Es cuestión de utilizar la tecnología apropiada. Por ejemplo, tomando una bacteria en su ambiente de forma que no se acople al entorno, si se la somete al experimento de la doble rendija[106], entonces no hay razón para que no se produzca una superposición cuántica[107].

Las evidencias experimentales a favor de una especificidad cuántica para el origen de la vida se apoyan en la exploración de las relaciones entre los mecanismos cuánticos de la biología, la presencia del fenómeno cuántico en algunas situaciones en biología y las novedosas propuestas que la computación cuántica ofrece al problema de la vida artificial.

Existen procesos específicamente cuánticos en la biología actual[108]:

1. Mutaciones cuánticas causadas espontáneamente por la alteración[109] de la estructura de las bases de los nucleótidos debido al efecto túnel protónico[110].
2. El efecto túnel facilita de manera importante la velocidad catalítica de algunas enzimas, aumentando considerablemente la eficiencia.
3. El origen del código genético muestra evidencias de optimización debidas a algoritmos cuánticos, como por ejemplo, el algoritmo de Grover[111].
4. En el rango de longitudes de los aminoácidos, en torno al nanómetro ($10^{-9}m$),

   pueden tener lugar alteraciones de la energía entre los monómeros que constituyen las caras de las superficie de las membranas celulares. Dichas alteraciones se deben a las variaciones de la energía cuántica del vacío provocadas por el efecto Casimir[112]
5. Efectos cuánticos en nano estructuras. Por ejemplo, no se sabe con certeza la naturaleza, cuántica o clásica, del límite que presenta la velocidad de ensamblaje de pares de nucleótidos que realiza la polimerasa[113]. Pero parece que la cota más plausible no se puede obtener por cálculos de tipo termodinámico y, tampoco, está determinada por la disponibilidad de

---

[103] Ver punto 4.3.3. dedicado a la medida en Mecánica Cuántica
[104] "Es difícil creer que la naturaleza no haya hecho un uso más inteligente de los efectos de la mecánica cuántica" ABBOTT, DAVIES Y PATI, op.cit. 355.
[105] Ver apéndice IV. Efecto Zenón.
[106] Ver experimento de la doble rendija p.45
[107] ABBOTT, DAVIES Y PATI, op.cit. p. 362.
[108] DAVIES, Paul C. W. "Does quantum mechanics play a non-trivial role in life?" *Biosystems*. 78, 2004, pp. 69-79.
[109] Está relacionado con el enlazamiento incorrecto del par.
[110] Ver Apéndice II.
[111] Ver p. 63
[112] Ver apéndice V.
[113] Enzima que utiliza energía en forma de ATP y como materia los nucleótidos se encarga de enlazar los pares de bases del DNA.

6. nucleótidos, sino que depende de la configuración geométrica y viene acotada por la incertidumbre cuántica de sincronización[114].
7. Aplicaciones de la ingeniería cuántica de la información como los juegos cuánticos donde surge el orden del desorden o los autómatas celulares.
8. Las muestras de fenómenos cuánticos que persisten actualmente en biología, por ejemplo en la fotosíntesis[115], son reliquias que evidencian la existencia anterior de un mundo completamente cuántico.

Tal vez, para aclarar si cabe aún más su conjetura, los autores apuntan que, como se ha afirmado, en la información biológica[116] está la clave para demostrar la "no trivialidad" o especificidad de la mecánica cuántica a la hora de abordar el origen de la vida. Postulan que además de su carácter clásico,[117] la información en biología tiene una propiedad cuántica[118], es decir, existen funciones biológicas y procesos de información en biología que no son clásicos sino cuánticos[119] y, por lo tanto, están codificados en qubits[120]. Por eso, eligen una definición de vida que prime la replicación: "The philosophical position that underpins my hypothesis is that the secret of life lies not with its complexity per se, still less with the stuff of which it is composed, but with its remarkable information processing and the replication abilities."[121]

Así pues, para sostener la propuesta cuántica, los autores adoptan una definición operativa de vida en la que todos los organismos vivos serían procesadores de información:

"… They store genetic data base and replicate it, with occasional errors, thus providing the basis for natural selection."[122]

Utilizan la analogía de von Neumann[123], que postula un autómata compuesto por un *hardware* proteico, encargado de realizar las funciones metabólicas, y un *software* que incorpora la información necesaria en forma de cadenas de nucleótidos. Los distintos tipos RNA lanzan un puente entre el *hardware* y el *software* para dotar a la célula de todas sus características funcionales y estructurales.

Para los autores de la propuesta cuántica del origen de la vida, las teorías biogénicas actuales describen la transición desde lo inerte hacia lo viviente sólo en términos del *hardware*. Proponen un cambio conceptual profundo, piensan que la clave para explicar la biogénesis se encuentra en el procesamiento de la información (*software*).

La cuestión determinante respondería a cómo se formó mediante síntesis abiótica la primera estructura autorreplicativa y cuál sería el papel que jugaría la mecánica cuántica en esa estructura. Si se recuerda que, uno de los grandes problemas del modelo del mundo del RNA de Eigen para explicar la emergencia de la autorreplicación era la baja probabilidad de ocurrencia natural, entonces se entenderá la importancia que puede tener la computación cuántica para disminuir el tiempo de exploración de los enlaces biológicamente interesantes[124]. Como se desprende de lo dicho, la naturaleza cuántica de ese "Primer Replicador" posibilitaría la vida, cuya emergencia destruiría los efectos cuánticos en casi todos los procesos salvo en determinadas excepciones que serían las observadas hoy en día. Por lo tanto, para continuar con el proyecto de investigación, es necesario describir la naturaleza de ese primer replicador. En este punto existen diferentes candidatos cuánticos.

---

[114] MESSIAH, Albert. *Mecánica cuántica: Tomo 1.* [Traducción por Carmen de Azcarate y Jaime Tortella]. Madrid: Tecnos, 1983.
[115] ABBOTT, Derek; DAVIES, Paul C. W. y PATI, Arun K. (Eds.). *Quantum aspects of life.* London: Imperial College Press, 2008, pp. 51-77.
[116] Ver apartado 3 sobre Biología e Información, p. 35
[117] Idem.
[118] ABBOTT, Derek; DAVIES, Paul C. W. y PATI, Arun K. (Eds.). *Quantum aspects of life.* London: Imperial College Press, 2008, p.358.
[119] Ibidem.
[120] Ver apartado 4.4.2. sobre Información cuántica.
[121] "La posición filosófica que subyace a mi hipótesis es que el secreto de la vida no se esconde en su complejidad per se y mucho menos en los elementos de los que se compone, si no que se encuentra en la gran capacidad de la propia vida para procesar información y sus habilidades de replicación". ABBOTT, DAVIES Y PATI, op. cit. p.4.
[122] "[los organismos vivos] guardan información genética y la replican, eso sí con errores ocasionales. Proveyendo de esta manera la base para la selección natural". ABBOTT, DAVIES Y PATI, op. cit. p.5.
[123] DYSON, Freeman J. *Los orígenes de la vida.* Traducción de Ana Grandal. Madrid: Cambridge University Press, 1999.

### 5.2.1  La Q-vida

Para Paul C. W. Davies, en la transición desde la materia inerte a la vida intermediaría un proceso cuántico. El candidato debería ser un sistema físico capaz de replicarse con gran precisión en tiempos suficientemente cortos. Tal vez usaría la orientación del espín electrónico o atómico y, por lo tanto, la información estaría codificada en binario. Aventura que podría tratarse de algún tipo de dispositivo de materia condensada a baja temperatura que podría encontrarse con facilidad en el espacio. A ese primer replicador lo denomina Q-vida. Presenta como ejemplo un caso simple para ilustrar la naturaleza de la Q-vida:

Considera[125] un conjunto de espines atómicos en cierta configuración $A \mid \uparrow\uparrow\uparrow\downarrow\downarrow\uparrow\uparrow\uparrow\downarrow\downarrow\uparrow\uparrow\uparrow\rangle$, que interacciona con otra muestra $B \mid \uparrow\downarrow\downarrow\downarrow\downarrow\uparrow\downarrow\uparrow\downarrow\uparrow\uparrow\uparrow\uparrow\rangle$ en la interacción, la evolución del estado o replicación de $AB \rightarrow AA$ :

$$\mid \uparrow\uparrow\uparrow\downarrow\downarrow\uparrow\uparrow\uparrow\downarrow\downarrow\uparrow\uparrow\uparrow\rangle \mid \uparrow\downarrow\downarrow\downarrow\downarrow\uparrow\downarrow\uparrow\downarrow\uparrow\uparrow\uparrow\uparrow\rangle \rightarrow \mid \uparrow\uparrow\uparrow\downarrow\downarrow\uparrow\uparrow\uparrow\downarrow\downarrow\uparrow\uparrow\uparrow\rangle \mid \uparrow\uparrow\uparrow\downarrow\downarrow\uparrow\uparrow\uparrow\downarrow\downarrow\uparrow\uparrow\uparrow\rangle$$

Se ha borrado la información en B, ya que el proceso es asimétrico e irreversible y va acompañado de un incremento de entropía porque requiere energía para cambiar uno a uno el espín de B, proceso que se realiza a gran velocidad.

Otro ejemplo más realista sería que el estado $A$ creara un estado complementario en el que cada espín fuera el opuesto $A = \mid \uparrow\uparrow\uparrow\downarrow\downarrow\uparrow\uparrow\uparrow\downarrow\downarrow\uparrow\uparrow\uparrow\rangle \; A_c = \mid \downarrow\downarrow\downarrow\uparrow\uparrow\downarrow\downarrow\downarrow\uparrow\uparrow\downarrow\downarrow\downarrow\rangle$ como los pares de bases conjugadas. Este modelo no tendría por qué contar con la interacción entre espines vecinos, situación que le aportaría más capacidad algorítmica. A estas estructuras las denomina Q-replicasa. Una vez que las dos estructuras se ponen en interacción, el salto de los espines, es decir, la replicación, es del orden de $10^{-15}s$ , doce órdenes de magnitud más rápida que la replicación natural (100 bases/segundo). Esta Q-replicasa utilizaría moléculas orgánicas para almacenar memoria al modo del *hardware* de un ordenador. Ahora el procesamiento cuántico es más rápido que el DNA o RNA pero necesita un aporte mayor de energía. De esta forma la molécula hubiera desarrollada la vida por sí misma, donde la pérdida de velocidad se compensaría por el incremento en complejidad, versatilidad y estabilidad de las moléculas, permitiendo que la vida invada todos los entornos con Q-vida. Estima que los límites de la replicación están acotados por el principio de incertidumbre y por la coreografía del sistema de replicación. Como al final el proceso duplicaría bits y no qubits, esto implica que no emergerían problemas debido a la decoherencia del sistema.

Otra posibilidad sería que las moléculas pudieran utilizar el poder computacional de la superposición cuántica para explorar simultáneamente todos los posibles enlaces con las moléculas de su alrededor. Tal vez, a través de la exploración cuántica de una estructura de decisión arborescente cuya raíz sea cualquier estado inicial y las ramas cualquier posibilidad en los estados de configuración del sistema. Las transiciones entre ramas se realizarían a gran velocidad mediante efectos cuánticos como el efecto túnel[126].

---

[124] Ver apartado 4.4.2. sobre Información Cuántica. Recuérdese que los fenómenos cuánticos del entrelazamiento y la superposición de estados, nos aumentan de forma inusitada la velocidad de procesamiento de la información.
[125] Punto por punto extractado de ABBOTT, Derek; DAVIES, Paul C. W. y PATI, Arun K. (Eds.). *Quantum aspects of life.* London: Imperial College Press, 2008, p. 8.

Esto acortaría tremendamente el tiempo estocástico que requiere el ensamblaje natural para conseguir moléculas auto replicativas. Ahora bien se debe explicar por qué una vez utilizada la superposición, el sistema colapsa en una configuración de molécula auto replicativa. La respuesta es que la interacción con el entorno es la causa del colapso y, por lo tanto, el entorno seleccionaría la configuración de las moléculas auto replicativas. Tal vez, la selección se produzca por efecto Zenón inverso[127] lo que significa que el ambiente favorece la vida sobre la no vida, por lo que emerge un tipo de teleología.

Al contrario de lo que ocurre con el DNA, en el que la arquitectura replicadora de las moléculas es la portadora de la información, en el caso cuántico, una vez realizada la replicación, la estructura de la información no es física. El autor ilumina esta afirmación con el ejemplo de los autómatas celulares cuánticos: "A cellular automaton consists of an infinite array of identical cells, the states of which are simultaneously updated in discrete time steps according to a deterministic role"[128].

Bajo ciertas condiciones de contorno, estos autómatas muestran comportamientos complejos. La clase de autómata más avanzada es la que muestra auto organización desde un estado estocástico. En general la información se pierde cuando se produce la transición entre estados. Sirva como ejemplo el modelo bidimensional cuadrupolar que requiere cuatro ítems: la dimensión de la muestra, el número de estados posibles para cada célula, el número de células finito que acotan cada unidad celular y las reglas de transición del autómata.[129] Por supuesto, la computación y algoritmia de las simulaciones es completamente clásica aunque actualmente se proponen modelos semi cuánticos. Estos proyectos de autómatas celulares cuánticos explotan la idea de que en ausencia de medida el sistema de células "vivo" está formado por la superposición de los estados de las células vecinas, tanto "vivas" como "muertas", todas representadas por osciladores clásicos con un determinado periodo, amplitudes entre cero y uno y fase variable. Este sistema muestra propiedades cuánticas básicas como la interferencia, descubre nuevas características de la complejidad del sistema y formula de manera más simplificada computacionalmente las propiedades complejas de los modelos clásicos.

Actualmente, la simulación computacional clásica, como el programa Tierra[130], parece mostrar la incapacidad de la formación espontánea de estructuras auto replicativas biológicas, aunque es conocida la facilidad para infectar la Red que presentan algunos programas autorreplicativos generados artificialmente (virus informáticos).

### 5.2.2 Las bibliotecas cuánticas combinatorias

El bioquímico John Joe Mcfadden y el físico Jim Al-Khalili[131] presentan otra hipótesis para explicar el origen de la vida. Están de acuerdo en que las estructuras vivas más simples son muy complejas. Éste es el caso del micoplasma cuyo tamaño genómico es de medio millón de pares de bases. Incluso las estructuras biológicas como los virus o la *Neoarchaeum equitans* son demasiado complejas para haber surgido espontáneamente de la sopa abiótica primordial. Estos dos investigadores conjeturan la necesidad de tener en cuenta estructuras más pequeñas que puedan realizar las funciones de almacenar la información y catalizar las reacciones bioquímicas. Se adhieren al paradigma biológico del mundo del RNA: el primer "gen" y la primera molécula catalizadora podría haber sido un tipo de RNA capaz de catalizar la formación de moléculas de RNA con idéntica secuencia. Estas moléculas auto replicantes sufrirían variaciones aleatorias y esas variantes de RNA entrarían en "competencia" por los nucleótidos de tal suerte que las moléculas más eficaces saldrían potenciadas. De la misma manera, estas moléculas podrían catalizar la concentración de aminoácidos para formar péptidos. El tándem RNA-péptido se sometería a la evolución molecular descrita de forma que algunas moléculas se especializarían en la conservación de la información (DNA), mientras que el RNA daría paso a las proteínas que con el tiempo desempeñarían eficazmente la función de catalizar.

---

[126] Ver apéndice III.

[127] Ver Apéndice IV.

[128] "Los autómatas clásicos son colecciones de células idénticas, dispuestas en series simples, cuyo estado se encuentra simultáneamente actualizado en estados discretos de acuerdo a reglas deterministas" ABBOTT, Derek; DAVIES, Paul C. W. y PATI, Arun K. (Eds.). *Quantum aspects of life.* London: Imperial College Press, 2008, p. 233.

[129] Ibidem.

[130] ABBOTT, Derek; DAVIES, Paul C. W. y PATI, Arun K. (Eds.). *Quantum aspects of life.* London: Imperial College Press, 2008, p.13.

[131] Ibidem, p.37.

Como se ha visto anteriormente, uno de los candidatos a primer replicador sería el ribosoma: una biomolécula muy compleja y difícil de sintetizar cuya estructura mínima auto replicativa consiste en 165 bases que suponen diez veces el tamaño máximo sintetizado en laboratorio. Es justamente esta baja probabilidad de formación espontánea mediante procesos aleatorios en la sopa primordial la que también sugiere a Mcfadden y Al-Khalili la necesidad de la gran capacidad de computación de la mecánica cuántica: "The problem is a search problem. The self-replicator is likely to be only one or few structures in a vast space of possible structures. The problem is that random search (essentially thermodynamic processes) is far too inefficient to find a self-replicator in any feasible period of time."[132]

Es decir, coinciden con la opinión de Davies y otros investigadores que enfocan el problema de la vida desde el punto de vista del almacenaje y tratamiento de la información. En palabras de Davies: "I am defining 'life' as a certain special state of low probability. Quantum mechanics enable the space of possibilities to be much more efficiently explored than a stochastic classical system"[133]

Para aumentar la velocidad de computación, los autores proponen un modelo de proto-autorreplicador que va más allá de ser una estructura que mediante mutaciones y exploración de su entorno químico consiga la estructura correcta del RNA primigenio o de los péptidos, ya que este proceso requeriría de cientos de millones de años para conseguir la emergencia de la primera estructura auto replicativa.

Mcfadden y Al-Khalili suponen que en las gotas del caldo primordial existían moléculas orgánicas suficientemente largas pero que tenían ordenados sus monómeros de forma incorrecta en el espacio secuencial. Esas estructuras moleculares podrían mutar debido a reacciones químicas. Encontrar la configuración correcta para la auto replicación en un tiempo razonable sería muy improbable, recuérdese que el tiempo de ensamblaje y ruptura de enlaces es muy lento y, además, el número de estructuras moleculares disponibles es limitado. Aunque se dispusiera de una biblioteca de estructuras dinámicas combinatorias, es decir, de una especie de "vademécum" de estructuras que pudiera formar todas las combinaciones de estructuras posibles, la formación química de macro moléculas sería demasiado lenta para evitar la degradación por procesos termodinámicos. Pero si el mecanismo de exploración de enlaces no es solamente clásico sino que es factible la exploración cuántica, entonces este "atajo cuántico" podría acertar con la estructura del protorreplicador en un tiempo razonable.

---

[132] "El problema es un problema de búsqueda. Es probable que el autorreplicador sea una o unas pocas estructuras dentro de un espacio inmenso de posibles estructuras. El problema es que la búsqueda aleatoria (esencialmente procesos termodinámicos) es demasiado ineficiente para poder encontrar un autorreplicador en un periodo de tiempo razonable". ABBOTT, Derek; DAVIES, Paul C. W. y PATI, Arun K. (Eds.). *Quantum aspects of life.* London: Imperial College Press, 2008, p.39.

[133] "Defino 'vida' como una especie de estado de baja probabilidad. La mecánica cuántica permite que el espacio de posibilidades pueda ser explorado con mayor eficacia que si se usara un sistema estocástico clásico". Ibidem, p.11

El mecanismo cuántico de exploración es de nuevo el efecto túnel que en química se conoce como *tautomerización*. En la tautomerización, el movimiento de los protones mediante efecto túnel altera las bases de los nucleótidos de tal suerte que quedan compuestas de una mezcla de formas en las que la posición de los protones cambia alternativamente de una base a la otra. Así pues, cada base molecular existe como superposición de varias formas tautoméricas ligadas mediante este efecto túnel protónico[134].

De nuevo la superposición cuántica se propone como solución al problema: las bases no serían $|1>$ (i.e. enol) o $|0>$ (etol) sino la superposición $a|01> \pm b|10>$. De esta forma se pueden crear conjuntos de macromoléculas "virtuales" que sean superposición de estados de moléculas pequeñas. En estas macromoléculas se forman unas colecciones inmensas, simultáneas y alternativas de posibles enlaces gracias al efecto túnel protónico[135].

"We now return to our primordial pool and imagine it to be a quantum dynamic combinatorial library with many molecules and a single compound that can each exist in a quantum superposition of tautomeric states simultaneously […]. If such a quantum superposition of all possible states in the combinatorial library can be built up before the onset of the coherence then this becomes an extremely efficient way of searching for the correct state: that of a simple replicator"[136].

Dos son los problemas principales de esta propuesta:

- El mantenimiento de la coherencia de la superposición de estados cuánticos, pues la interacción del estado con el entorno destruye con suma facilidad el delicado estado de superposición[137].
- La teleología subyacente al problema del mecanismo que "elige" al replicador molecular como estructura final.

Mcfadden y Al-Khalili solucionan el segundo problema mediante el conocido efecto cuántico selectivo de la medida denominado "acto irreversible de amplificación" por el cual, una vez que el escrutinio del estado molecular pasa por el del replicador éste se hace macroscópicamente distinguible debido a su fuerte interacción con el entorno, cuando se produce la transición hacia la replicación química, que induce la decoherencia cuántica de forma fulminante convirtiéndole en replicador "clásico".

No obstante queda pendiente cómo mantener la coherencia del estado de superposición cuántico el mayor tiempo posible para que sea factible la exploración de los numerosos estados cuánticos posibles y disponga del tiempo suficiente a escala biológica para encontrar el estado del replicador antes de ser destruida.

Es posible evitar el colapso del estado en superposición mediante varias técnicas: bajando la temperatura de la muestra molecular, manteniendo aislada la muestra del entorno – mediante apantallamiento molecular- o mediante el efecto conocido como "espacios libres de decoherencia". Paradójicamente, el fuerte acoplamiento con el entorno de ciertos grados de libertad del sistema provoca de forma sorprendente la "congelación" del resto de los grados de libertad[138], permitiendo el mantenimiento de la superposición coherente e incluso la persistencia de la cualidad cuántica del enredo o entrelazamiento.

---

[134] Ver apéndice III.
[135] Ver apéndice III
[136] "Podemos imaginar nuestro caldo primordial cuya biblioteca combinatoria dinámica se compone de muchas moléculas de un único componente que puede existir como superposición cuántica de muchos estados tautoméricos simultáneos […] Si se puede forma tal superposición cuántica de todos los posibles estados en la biblioteca combinatoria antes de que tenga lugar la decoherencia, entonces ésta es la forma más eficiente de buscar el estado correcto: el de un replicador simple". ABBOTT, Derek; DAVIES, Paul C. W. y PATI, Arun K. (Eds.). *Quantum aspects of life.* London: Imperial College Press, 2008, p.42.
[137] Ver apartado 4.3.3. para la medida y el colapso de estados.

## 5.3 Algoritmos cuánticos en el origen de los lenguajes genéticos

Las investigaciones de Apoorva D. Patel se centran en la comprensión del origen de los lenguajes genéticos. Sus características son ideales para construir una teoría de la información porque son lenguajes universales: cinco nucleótidos usados como material genético, y veinte aminoácidos para la construcción de las proteínas. Además la información genética se codifica con una optimización alta entre el espacio y el límite de compresión y los cambios locales están representados por la ocurrencia estadística de mutaciones. En definitiva, piensa que los lenguajes del DNA y las proteínas siguen criterios de eficiencia y propone: "[…] look at life as an exercise in information theory, and extend the analisys as far as posible".[139]

Patel apunta que no es lo mismo lo que se entiende por eficiencia en biología que en teoría de computación porque en computación debe existir, como se ha visto, un equilibrio entre espacio y tiempo. En el caso de la biología, el tiempo requerido para cualquier proceso es de vital importancia, mientras que el espacio y la materia se utilizan en grandes cantidades para que pueda tener lugar la evolución.

Su hipótesis de trabajo es que el lenguaje genético se desarrolló gracias al método de prueba y error que, una vez conseguido el grado de optimización actual, se estabilizó gracias a la selección natural. La optimización se alcanzó en el equilibrio entre el tiempo disponible de exploración de las distintas configuraciones y el tiempo de cambio o transición. Para conseguir esta optimización cree que es preciso: un lenguaje digital, que utiliza el número mínimo de instrucciones y soporta alta tolerancia a los errores; materiales simples, esto facilita su disponibilidad natural; respuestas automáticas del *hardware*, evitando la necesidad de traducción; y operaciones rápidas, que permitan la estabilidad frente a los procesos de degeneración. De esta forma se podrá obtener una alta densidad de empaquetamiento y velocidad.

No es suficiente la física clásica para describir todos los enlaces que se presentan en las moléculas del material genético o de las proteínas, debido a que el proceso de identificación de los enlaces de hidrógeno es muy sensible a las correlaciones cuánticas a causa del efecto túnel protónico[140] apuntado más arriba. Por lo tanto, conjetura la estructura atómica como el *hardware* del sistema, que restringe las operaciones de implementación física, y el algoritmo cuántico de Grover[141] como el *software[142]*, que indica cuál es la tarea a desarrollar, en este caso la búsqueda en una inmensa base de datos con la mejor eficiencia[143]. Evidentemente, vuelve a surgir como problema la posibilidad de mantener la coherencia entre las configuraciones de transición. No obstante propone que ese tiempo de coherencia sea tal que no supere el tiempo necesario para identificar los enlaces y no sobrepase el tiempo en el que se alcanza el equilibrio.

---

[138] Efecto Zenón. Ver apartado 4.3.3. para el problema de la medida en cuántica y también el Apéndice IV.
[139] "[…] mirar la vida como un ejercicio de teoría de la información y extender el análisis tan lejos como sea posible." ABBOTT, Derek; DAVIES, Paul C. W. y PATI, Arun K. (Eds.). *Quantum aspects of life*. London: Imperial College Press, 2008, p.187.
[140] Ver apéndice III
[141] p.63
[142] Sus estados inicial y final son clásicos, pues se requiere una medida.
[143] Ver nota 96.

## CONCLUSIONES

La naturaleza cuántica del mundo es el principio del que surge la vida. La replicación es la característica definitoria de la vida y ésta se reduce a la capacidad de mantener y procesar información. Por lo tanto, el origen de la vida depende de que las moléculas abióticas del caldo prebiótico primigenio anterior al mundo del RNA puedan auto ensamblarse para convertirse en estructuras auto replicativas. Gracias a las propiedades no triviales de la mecánica cuántica, como la superposición de estados, aumenta exponencialmente la capacidad de exploración de enlaces entre las moléculas del caldo primigenio y, por lo tanto, se eleva considerablemente la probabilidad de que las moléculas prebióticas encontraran el protorreplicador del que procede el material genético de los sistemas biológicos. Los problemas más difíciles que afronta la teoría son:

1. Es sabido que, actualmente, no se puede mantener la coherencia en las escalas de tiempo de los sistemas biológicos. Entonces: ¿cómo se pudo mantener la coherencia cuántica en el mundo prebiótico dominado por la termodinámica?
2. ¿Cómo se puede evitar la teleología subyacente a la elección de la estructura protorreplicativa entre todas las posibles?

Ahora resta contestar algunos de los interrogantes que surgen al afrontar el problema del origen de la vida

▪ El mundo del RNA sigue siendo la teoría más aceptada de la transición entre el mundo prebiótico y la vida aunque uno de sus puntos débiles es la baja probabilidad de ensamblaje espontáneo de los elementos esenciales para el metabolismo y la replicación. Se recordará que la eficiencia y las posibles catástrofes de los errores suponen una dificultad que parece insalvable. En lo que se refiere a estas cuestiones, la propuesta cuántica se postula como la mejor forma de eludir la baja probabilidad de formación espontánea debido a la potencia de exploración de enlaces que muestra la mecánica cuántica gracias a las propiedades de superposición y enredo. Además, la eficiencia y fidelidad que muestran los algoritmos cuánticos como los de Grover, es superior a las posibilidades que propone la teoría clásica del enlace. Por lo tanto, a primera vista parece que la conjetura cuántica sobre el origen de la vida aumenta la credibilidad de la existencia de un hipotético mundo del RNA. Ahora bien, la hipótesis cuántica guarda silencio ante problemas como la catástrofe del error y no deja claro cómo mantener la coherencia en la superposición de estados para poder llevar a efectos la exploración de los enlaces. En este orden de cosas, sostienen la capacidad de encontrar la configuración correcta en una conjetura, ya apuntada, denominada "acto irreversible de amplificación"[144]. Por último, para muchos científicos, no se necesita escudriñar las propiedades exóticas del mundo cuántico si lo que se pretende es dar cuenta de los enlaces químicos.

▪ Los autores pretenden defender la contingencia del origen de la vida utilizando la gran capacidad computacional de la mecánica cuántica. Si la exploración de las posibles configuraciones moleculares de los protorreplicadores se realiza cuánticamente, la improbabilidad clásica de que la vida surja espontáneamente del caldo primordial se transforma en necesidad.

A mi entender, defienden un determinismo procedente de lo que entienden por "principio de la vida" que interpretan como ley natural al modo de la cosmología actual. Algunos apoyan sus argumentos en el denominado principio antrópico fuerte en el que las constantes universales están ajustadas para que surja la vida. Esto parece que implica que la vida está ligada a la cosmología cuántica[145] y a la teoría de supercuerdas. Como consecuencia de la teoría sobre la medida de Everett[146], conjeturan que la vida es una de las posibilidades de los múltiples universos y surge en nuestro mundo debido a la configuración justa de las constantes universales. El asunto a nivel experimental es saber de qué forma podemos conocer las constantes de otros mundos, para poder afirmar que la probabilidad de la aparición de ciertas constantes determinan la existencia de la vida. Aunque, más bien parece que simplemente la probabilidad de la vida es la unidad debido a que en nuestro mundo existen seres vivos.

---

[144] Ver apéndice III.
[145] Davies, P.C.W. "Multiverse cosmological model". *Modern Physics Letters A.*, 19, (10), 2004, pp. 727-734.

En biología, el principio antrópico fuerte conduce al programa genético, que los autores de la conjetura cuántica reducen a información genética.

Como se sabe, la causación en biología[147] sigue mostrando una influencia importante del marco aristotélico[148]. La orientación desde el nivel físico-químico al biológico sostenida por los autores es típica del reduccionismo pues tanto las propiedades como las leyes del sistema en el nivel biológico son determinadas por las propiedades y las leyes de sus componentes. Pero, al ignorar la orientación descendente, es decir, desde el nivel jerárquico superior de los biosistemas al inferior de las moléculas que componen la célula, no consideran la importancia del entorno en la replicación y, por lo tanto, no queda claro cómo los protorreplicadores se convierten en material genético. Davies propone la Q-vida, como el protorreplicador que cumple funciones de procesado de la información, aunque queda reducido a las funciones de la denominada Q-replicasa a la que se le aplican los conceptos de azar y probabilidad. No queda claro, de qué forma este principio activo pasa a convertirse en la macromolécula pasiva del DNA, ya que la propuesta teórica del origen cuántico de la vida no es causal ya que la causa estricta no es estocástica[149], pues, aunque pueda modificar las propensiones no es la propensión misma.

Por otro lado, las evidencias sobre propiedades cuánticas de la biología que se presentan en la propuesta son, en el mejor de los casos, relaciones funcionales entre propiedades y leyes a distintos niveles. Por ejemplo, esto sucede cuando se afirma que el efecto túnel facilita de manera importante la velocidad catalítica. En otros casos, simplemente se presentan como evidencias meras conjeturas como, por ejemplo, defender que la optimización de los lenguajes genéticos es debida a la participación de algoritmos cuánticos.

En lo que respecta a las causas finales, una de las preocupaciones más importantes de los autores es evitar la teleología en la selección del replicador:

"There seems to be an unavoidable teleological component involved: the system somehow 'selects' life from the vastly greater numbers of states that are nonliving."[150]

Davies deja abierta la puerta a la posibilidad de una causa final. Sin embargo, Mcfadden y Khalili quitan relevancia a la presunta teleología:

"…an element of teleology is required; namely that the molecule must somehow know beforehand what is aiming for. We do not believe this is necessary (as we argue below)"[151].

Su argumento se vuelve a fundamentar en un hipotético "acto irreversible de amplificación" que requiere de justificación experimental en el nivel molecular.

---

[146] TORRETTI, Roberto. *The Philosophy of Physics.* Cambridge: University Press, 1999.
[147] En este punto se sigue a MAHNER, Martín y BUNGE, Mario. Fundamentos de biofilosofía. Méjico: Siglo XXI, 2000.
[148] ABBOTT, Derek; DAVIES, Paul C. W. y PATI, Arun K. (Eds.). *Quantum aspects of life.* London: Imperial College Press, 2008.
[149] Ibidem, p.55
[150] "Parece estar relacionado un componente teleológico inevitable: el sistema, de alguna forma 'selecciona' la vida de una gran número de estados inertes" ABBOTT, Derek; DAVIES, Paul C. W. y PATI, Arun K. (Eds.). *Quantum aspects of life.* London: Imperial College Press, 2008, p.11
[151] "… se requiere un elemento de teleología; concretamente, de alguna forma la molécula tiene que saber de ante mano su objetivo."Ibidem, p.42

▪ Una de las cuestiones centrales de la propuesta cuántica sobre el origen de la vida es la propia definición de vida que utilizan los autores. Por un lado, como se ha apuntado anteriormente, entienden la vida como un proceso:

"I'm defining 'life' as a certain special state of low probability."[152]

"The philosophical position that underpins my hypothesis is that the secret of life lies not with its complexity per se, still less with the stuff of which is composed, but with is remarkable information processing and replication abilities."[153]

"Apparently, all that is required for proto-life is the existence of physical systems that reproduce themselves with variations. As much as one might like to set other requirements for the origin of life, reproduction and variation seem to suffice"[154]

Los autores no definen la vida, realmente postulan la replicación como la única propiedad intrínseca de lo que ellos entienden por estar vivo, desentendiéndose de las demás cualidades que presenta la vida. Esta propiedad emergente a veces la presentan como fuerte, este es el caso que presenta la Q-vida:

"The starting point of my hypothesis is the existence of a quantum replicator, a quantum system that can copy information with few errors […] Quantum replicators certainly exist in Nature."[155]

Como se ha apuntado anteriormente, en todos los autores se aprecia una tendencia a la reificación: por ejemplo, toman como cosas las relaciones propuestas para las distintas bibliotecas combinatorias dinámicas. Esto se debe, tal vez, a la tendencia de los físicos a cosificar los constructos teóricos de la mecánica cuántica. Además, en los distintos artículos, es normal encontrar enunciados metafísicamente mal formados que se entienden como costumbres del lenguaje coloquial pero que llegan a ser fuente importante de confusión cuando se pretende aclarar el origen de la vida, por ejemplo: "The self-replicator is likely to be only one or few structures in a vast space of possible structures"[156]. Si el auto replicador es una cosa entonces no puede ser un constructo ficticio.

De la misma forma los autores al explicar los fenómenos utilizan mecanismos y procesos que olvidan que no poseen existencia fuera de las cosas confundiendo lo que es pura abstracción metodológica con la tesis ontológica que deja en segundo plano las cosas para primar los eventos o los procesos.

Por último, opino que la vida no es la propiedad de una sola molécula o de todas y cada una de las moléculas por separado, sino de un gran número de ellas con cierta organización y que, bajo las condiciones favorables del ambiente, han evolucionado de forma ascendente en el nivel de complejidad. La definición utilizada por los autores no tiene en cuenta que una vez encontrado el proto replicador, éste tiene que comportarse como un sistema clásico y por lo tanto está condicionado por el entorno y ligado a las propiedades del conjunto de moléculas. Es decir, puede no ser suficiente la estructura correcta para la replicación si el entorno es del estilo de las catástrofes del error. Por eso, una definición reduccionista como la propuesta, la vida es información, no puede dar cuenta de un fenómeno complejo como el origen de la vida.

---

[152] "Defino 'la vida' como un cierto estado de baja probabilidad". Ibidem, p.11
[153] "La posición filosófica que subyace a mi hipótesis es que el secreto de la vida no se esconde en su complejidad per se y mucho menos en los elementos de los que se compone, si no que se encuentra en la gran capacidad de la propia vida para procesar información y sus habilidades de replicación". Ibidem, p.4
[154] "Aparentemente lo único que se requiere para la protovida es la existencia de sistemas físicos que se auto repliquen con variaciones. Por mucho que uno quiera exigir otros requerimientos al origen de la vida, parece que con la replicación y la variación es suficiente" Ibidem, p.27
[155] "El punto de partida de mi hipótesis es la existencia de un replicador cuántico capaz de copiar información con pocos errores […]. Ciertamente existen replicadores cuánticos en la naturaleza". ABBOTT, Derek; DAVIES, Paul C. W. y PATI, Arun K. (Eds.). *Quantum aspects of life.* London: Imperial College Press, 2008, p.7.
[156] "Es probable que el autorreplicador sea una o unas pocas estructuras dentro de un espacio inmenso de posibles estructuras". ABBOTT, DAVIES y PATI, op. cit., p.39.

▪ Los autores defienden un concepto unitario de la vida basado en la conjetura de que lo vivo es equivalente a lo que tiene capacidad de replicarse. No entran en valoraciones sobre la posibilidad de un doble origen de la vida metabólico y replicador. A pesar de lo expuesto, presentan un mecanicismo de "doble vía": por un lado es evidente que defienden el fisicoquimicalismo, donde los organismos son sistemas fisicoquímicos complejos sin leyes propias y, por otro, presentan una renovada versión del maquinismo en la línea de la tradición cartesiana y de La Mettrie: la vida y los organismos se presentan como máquinas de computación distinguiendo el *hardware* del *software*. Esta descripción cuadra con la idea de autómatas replicantes dentro del proyecto de vida artificial y en algunos casos con la idea de que "la biología no sólo es como la ingeniería; es ingeniería".[157]

Esta última cuestión enlaza con la propuesta reduccionista de los autores. Paul Davies se declara[158] en la línea de las ideas de Kauffman sobre la relevancia de la auto organización[159]: *"la auto organización es el proceso en el cual las interacciones locales entre los elementos de un sistema producen patrones emergentes de comportamiento sin que para ello sea necesario algún tipo de coerción o control externo. Estos patrones o comportamientos surgen en ausencia de un diseño o plan central y se consideran emergentes porque no pueden ser deducidos a partir del conocimiento total de los elementos de menor nivel ni de la naturaleza de las interacciones entre ellos"*. Propone que su modelo constituya una versión cuántica del concepto que Kauffman defiende como sistema auto catalítico de moléculas.

Las ideas de Davies, Lloyd, Mcfadden y Al-Khalili y las procedentes de los sistemas de Q-vida computacional se adaptan a la propuesta de que:

"…en ausencia de la selección el sistema caerá en propiedades promedio bien definidas […] los patrones de inferencia buscan explicar propiedades biológicas localizándolas como elementos genéricos de un ensamblaje apropiado a través del cual la evolución se mueve."[160]

Así pues podemos decir que el problema de la reducción desde el punto de vista epistémico para nuestros autores tiene su precedente y origen en la propuesta de la síntesis *kauffmiana*. Obviamente, las fórmulas puente de la reducción o principios puente residen en las hipótesis cuánticas que defienden. La reducción conceptual y proposicional está presente en todos los artículos, especialmente en el de Davies, que enuncia definiciones reductivas como la Q-vida que parecen apuntar a la intención de reducir la teoría molecular a la teoría cuántica aunque el intento no deja de presentar soluciones reductivas de tipo débil pues utiliza, además de la citada reducción de conceptos, hipótesis auxiliares como: "The direction of information flow is bottom up"[161]

Opino que no se consigue una reducción de la biología a las leyes de la física, ni los autores aportan un argumento nuevo al problema de la reducción con este programa de investigación, pues utilizan argumentos analógicos que carecen de peso como la irracionalidad de que la vida se interprete clásicamente, siendo ésta producto de la actividad molecular que cae en el dominio de la mecánica cuántica. Opino que ni siquiera la física es reducible de forma fuerte a una teoría básica como la de la mecánica cuántica, ya que incluso ésta necesita de conceptos clásicos (masa, tiempo…), así como de hipótesis sobre el mundo macrofísico.[162]

---

[157] MAHNER, Martín y BUNGE, Mario. *Fundamentos de biofilosofía.* Méjico: Siglo XXI, 2000, p.166.

[158] ABBOTT, Derek; DAVIES, Paul C. W. y PATI, Arun K. (Eds.). *Quantum aspects of life.* London: Imperial College Press, 2008, p.12.

[159] ANDERSON, C. Apud. PÉREZ MARTÍNEZ, Alfredo. *La obra de Stuart Kauffman: aportaciones a la biología del siglo XXI e implicaciones filosóficas*. [Tesis T.F.M (Trabajo fin de Máster). Universidad Complutense de Madrid, 2005], p.131.

[160] Apud PÉREZ MARTÍNEZ, Alfredo. *La obra de Stuart Kauffman: aportaciones a la biología del siglo XXI e implicaciones filosóficas*. [Tesis T.F.M (Trabajo fin de Máster). Universidad Complutense de Madrid, 2005], p.131.

[161] "La dirección que sigue el flujo de información es de 'arriba-abajo'". ABBOTT, Derek; DAVIES, Paul C. W. y PATI, Arun K. (Eds.). *Quantum aspects of life.* London: Imperial College Press, 2008, p.5.

Así mismo, se puede notar en los autores una tendencia al atomismo ontológico apoyada en el micro reduccionismo aunque, a mi entender, esto es sobre todo una estrategia de investigación.

Si seguimos las ideas de reducción de Rosenberg [163] el reduccionismo propuesto por los autores tiene fundamento metafísico y se caracteriza claramente por defender una metodología explicativa o subsunción ascendente, que estimo insuficiente para determinar el origen de la vida.

- Al afirmar[164] que la mecánica cuántica permitió la emergencia de la vida desde el nivel atómico sin la intermediación de la química compleja, parece que alguno de los autores defiende una reducción fuerte de la biología a la física y una emergencia de raíz epistemológica fundamentada en la propuesta cuántica. Pero, pienso que no se puede admitir una reducción fuerte y, por otro lado, la emergencia de la vida es, a mi parecer, ontológica. Mi argumento es que la propiedad de estar vivo no la tienen las partes de la célula. Los aminoácidos, las proteínas y las demás componentes de la bioquímica celular no son vida por más que sean necesarias. Por lo tanto, si las partes no tienen la propiedad de estar vivo entonces esta propiedad es ontológicamente emergente.

Para algunos autores, como Seth Lloyd[165], la emergencia de la complejidad del mundo orgánico puede tener una explicación plausible en términos cuánticos gracias a la discontinuidad y la probabilidad o aleatoriedad inherente a los fenómenos cuánticos (fluctuaciones cuánticas). Según Seth Lloyd[166], los diferentes niveles de complejidad como el de las reacciones químicas y la vida exploran las distintas posibilidades permitidas por las leyes físicas. La base de estos fenómenos surge de "accidentes mecanocuánticos" que son procesados por el poder computacional del universo. En palabras de Lloyd:

"… quantum mechanics then adds the variety that is the proverbial spice of life. The result is what we see around us."[167]

Parece que la emergencia de la vida es una propiedad inherente a la capacidad computacional del universo como un todo. Tal vez expresión de una visión renovada del Demiurgo.

Este tipo de conjeturas tratan la vida como el caso paradigmático de una forma dinámica de emergencia "débil"[168], pues, queda claro a lo largo del texto, que el problema a solventar es el de la baja probabilidad de un ensamblaje espontáneo de los replicadores, por lo tanto, las propiedades macroscópicas son impredecibles y no se pueden derivar salvo mediante la observación del proceso o de la simulación. No obstante, a mi parecer la emergencia es independiente de que se conozca o no el mecanismo por el que se consigue tal propiedad. Es decir, si una propiedad es totalmente nueva su emergencia es ontológica.

---

[162] MAHNER, Martín y BUNGE, Mario. *Fundamentos de biofilosofía.* Méjico: Siglo XXI, 2000, p.140.
[163] SARKAR, Sahotra y PLUTYNSKI, Anya (Eds.). *A Companion to the Phylosophy of Biology.* Oxford: Blackwell Publishing, 2007, p.550-568.
[164] ABBOTT, Derek; DAVIES, Paul C. W. y PATI, Arun K. (Eds.). *Quantum aspects of life.* London: Imperial College Press, 2008.
[165] ABBOTT, Derek; DAVIES, Paul C. W. y PATI, Arun K. (Eds.). *Quantum aspects of life.* London: Imperial College Press, 2008, p. 29.
[166] ABBOTT, Derek; DAVIES, Paul C. W. y PATI, Arun K. (Eds.). *Quantum aspects of life.* London: Imperial College Press, 2008, p. 29.
[167] "… la mecánica cuántica añade la variedad que es la sal de la vida. El resultado es lo que vemos a nuestro alrededor". Ibidem.
[168] BEDAU, Mark A. "What is life?" En: SARKAR, Sahotra y PLUTYNSKI, Anya. *A Companion to the Phylosophy of Biology.* Oxford: Blackwell Publishing, 2000, p.459.

En resumen, las propuestas estudiadas en este trabajo sobre una especificidad cuántica para entender el origen de la vida abren una nueva vía de investigación en lo que se refiere al problema del surgimiento de la replicación y a la explicación de la fidelidad y eficiencia del lenguaje genético. Las aplicaciones de la computación cuántica apuntan un futuro prometedor en el almacenamiento, procesamiento y encriptación de la información en biosistemas. Representan un camino nuevo a la miniaturización de los sistemas de computación y a la *biorobótica*.

Por otro lado, creo que no supondrá una novedad en cuanto a la explicación de la emergencia de la vida, mientras siga con la pretensión de una programa reduccionista fuerte incapaz de incorporar todas y cada una de las propiedades de la vida.

Creo que la biología del futuro puede beneficiarse enormemente si analiza la relevancia de la mecánica cuántica como puente entre la bioquímica y la biología molecular. La clave del éxito puede encontrarse en impulsar la interdisciplinaridad y en conjugar la visión sistémica con el reduccionismo débil, de estas dos estrategias de investigación surgirán las preguntas correctas.

# **BIBLIOGRAFÍA**

ABBOTT, Derek; DAVIES, Paul C. W. y PATI, Arun K. (Eds.). *Quantum aspects of life.* London: Imperial College Press, 2008.

BEAM Line: a periodical of particle physics [en línea]. Stanford: Stanford Linear Accelerator Center. Vol. 30 (2), 2000, 63 p. [Disponible versión electrónica http://www.slac.stanford.edu/pubs/beamline/pdf/00ii.pdf. Consultado por última vez 06/09/2010]

BERG, Jeremy M., STRYER, Lubert y TYMOCZKO, John L. *Bioquímica* [versión española por José M. Macarulla]. Barcelona: Reverté, 2003.

BOHM, David. *Quantum Theory*. New York: Dover Publications, 1989.

BOHR, Niels. "Quantum Mechanics and Physical Reality". *Nature,* 136, 1935, pp. 1025-1026.

BOHR, Niels. "The quantum postulate and the recent development of atomic theory". *Nature*, 121, 1928, pp. 580-591.

BOITEAU, Laurent y PASCAL, Robert. "Energy Sources, Self-organization, and the Origin of Life". *Origins of Life and Evolution of Biospheres,* 2010. [Disponible sólo versión electrónica http://www.springerlink.com/content/yu37121113850r06/. Consultado por última vez 05/09/2010].

BREUR, Hans. *Atlas de Química. Vol. 1. Química general e inorgánica; Vol. 2. Química orgánica y polímeros.* Versión española de Ricardo Serrano Roche y Mª Gloria Quintanilla López. Madrid: Alianza, 1988, 2 v.

CASSIDY, David, HOLTON, Gerald y RUTHERFORD, James. *Understanding Physics.* New York [et al.]: Springer, 2002.

CASTRODEZA, Carlos. *Los límites de la Historia Natural.* Tres Cantos, Madrid: Akal, 2003.

CIRAC SASTURÁIN, Juan Ignacio. "Quanta y computación". *Revista española de Física,* 14 (1), 2000, pp.48-52.

CHYBA, Christopher y HAND, Kevin P. "Astrobiology: the Study of the Living Universe". *Annual Review of Astronomy and Astrophysics,* 43, 2005, pp. 31-74.

CLELAND, Carol y CHYBA, Christopher. "Defining life". *Origins of Life and Evolution of Biospheres,* 32 (4), 2002, pp. 387-393.

COOPER, W. Grant. "Necessity of Quantum Coherence to Account for the Spectrum of Time-Dependent Mutations Exhibited by Bacteriophage T4". *Biochemical Genetics,* 47 (11-12), 2009, pp. 892-910.

*CUADERNOS de Ontología = Ontology Studies*. San Sebastián: Departamento de Filosofía, Universidad del País Vasco. 2001. Nº 1-2. ISSN 1576-2270.

CURTIS, Helena y BARNES, Sue. *Biología.* [Dir. Adriana Schnek y Graciela Flores]. Madrid: Médica Panamericana, 2001.

DAVIES, Paul C. W. "Does quantum mechanics play a non-trivial role in life?" *Biosystems.* 78, 2004, pp. 69-79.

Davies, Paul C. W. "Multiverse cosmological model". *Modern Physics Letters A.*, 19, (10), 2004.

DÍEZ, José y MOULINES, Carlos Ulises. *Fundamentos de filosofía de la ciencia.* Barcelona: Ariel, 1999.

DYSON, Freeman J. *Los orígenes de la vida.* Traducción de Ana Grandal. Madrid: Cambridge University Press, 1999.

EIGEN, Manfred et al. "The origin of genetic information". *Scientific American*, 4, (244), 1981, pp. 88-118.

EINSTEIN, Albert; PODOLSKY, Boris y ROSEN, Nathan. "Can quantum-mechanical description physical of reality be considered complete?" *Physical Review,* 47, 1935, pp. 777-780.

GALINDO TIXAIRE, Alberto. "Quanta e Información". *Revista española de física*, 14 (1), 2000, pp.30-46.

GILMOUR, Iain y SEPHTON, Mark A. (Eds.). *An introduction to Astrobiology.* Cambridge: University Press, 2004.

GLASER, Roland. *Biofísica.* [Traducción Félix Royo López y Félix Royo Longás]. Zaragoza: Acribia, 2003.

GONZÁLEZ RECIO, José Luis. *Átomos, almas y estrellas: estudios sobre la ciencia griega.* Villaviciosa de Odón, Madrid: Plaza y Valdes, 2007.

GONZÁLEZ RECIO, José Luis (ed.). *Philosophical essays on Physics and Biology.* Hildesheim, Zurich, New York: Georg Olms, 2009.

GRIFFITHS, David J. *Introduction to Quantum Mechanics.* New Jersey: Prentice Hall, 2004.

GROVER, Lov K. "Quantum Mechanics Helps in Searching for a Needle in a Haystack" *Physics Review Letters*, 79 (2), 1997, pp. 325-328.

HAYDON, Nathan, MCGLYNN, Shawn E. y ROBUS, Olin. "Speculation on Quantum Mechanics and the Operation of Life giving Catalysts". *Origins of Life and Evolution of Biospheres,* 2010. [Disponible sólo versión electrónica: http://www.springerlink.com/content/m156l241044p4h2j/. Consultado por última vez: 05/09/2010].

HEISENBERG, Werner. *The physical principles of the Quantum Theory.* Chicago: Dover Editions, 1930.

*INVESTIGACIÓN y ciencia.* 2003. Temas 31: Fenómenos cuánticos. ISSN 0210-136X.

JAKOSKY, Bruce. *Science, society, and the search for life in the Universe.* Tucson: University of Arizona Press, 2006.

KOOLMAN, Jan y RÖHM, Klaus-Heinrich. *Bioquímica: texto y atlas.* [Traducción de Lorenzo Facorro y Beatriz Medina de Thierre]. Madrid: Médica Panamericana, 2004.

LEHNINGER, Albert. *Principios de bioquímica.* David L. Nelson y Michael M. Cox (eds.). Barcelona: Omega, 2000.

LUIGI LUISI, Pier. *La vida emergente: de los orígenes químicos a la biología sintética.* Traducción de Ambrosio García Leal. Barcelona: Tusquets, 2010.

MAHNER, Martín y BUNGE, Mario. *Fundamentos de biofilosofía.* Méjico: Siglo XXI, 2000.

MAINZER, Klaus. *Thinking in Complexity: the Complex Dynamics of Matter, Mind and Mankind.* Berlin [et al.]: Springer, 1997.

MCFADDEN, Johnjoe. "A quantum mechanical model of adaptative mutation". *Biosystems,* 50 (3), 1999, pp. 203-211.

MESSIAH, Albert. *Mecánica cuántica: Tomo 1.* [Traducción por Carmen de Azcarate y Jaime Tortella]. Madrid: Tecnos, 1983.

MONOD, Jacques. *El azar y la necesidad: ensayo sobre la filosofía natural de la biología moderna.* [Traducción Francisco Ferrer Lerín, rev. por Antonio Cortés Tejedor]. Barcelona: Orbis, 1985.

MONTERO, Francisco y MORÁN, Federico. *Biofísica.* Madrid: Eudema, 1992.

NIELSEN, Michael y CHUANG, Isaac. *Quantum computation and quantum information*. Cambridge: University Press, 2000.

OPARÍN, Alexander. *El origen de la vida.* Madrid: Labor, 1977.

ORGEL, Leslie E. "Prebiotic Chemistry and the Origin of the RNA World". *Critical Reviews in Biochemistry and Molecular Biology,* 39, 2004, pp. 99-123.

PASSARGE, Eberhard. *Genética: texto y atlas.* [Traducción de Viviana Bumaschny, Karen Mikkelsen y Daniela Noaín]. Madrid: Médica Panamericana, 2009.

PATEL, Apoorva. "Quantum algorithms and the genetic code" [http://xxx.lanl.gov/abs/quant-ph/000207. Consultado por última vez 10/09/2010].

PATEL, Apoorva. "Testing quantum dynamics in genetic information processing". [http://xxx.lanl.gov/PS_cache/quant-p/0102034v2.pdf. Consultado por última vez el 10/09/2010].

PATEL, Apoorva. "Why genetic information processing could have a quantum basis". [http://xxx.lanl.gov/PS_cache/quant-p/0105001v2.pdf. Consultado por última vez el 10/09/2010].

PÉREZ MARTÍNEZ, Alfredo. *La obra de Stuart Kauffman: aportaciones a la biología del siglo XXI e implicaciones filosóficas*. [Tesis T.F.M (Trabajo fin de Máster). Universidad Complutense de Madrid, 2005].

PLATTNER, Helmut y HENTSCHEL, Joachim. *Manual de biología celular.* Traducción de Luis Serra, revisión y prólogo Mercé Dufort. Barcelona: Omega, 2001.

RASMUSSEN, Steen. "Bottom-Up will be more telling" *Nature*, 456, 2010, p. 422. [http://www.nature.com/nature/journal/v465/n7297/pdf/465422a.pdf]. [Consultado por última vez 04/09/2010].

RETAMOSA GRANADO, Joaquín, TEJERO CANTERO, Álvaro y RUIZ MÚZQUIZ, Pablo. *Introducción a la física cuántica.* [Disponible versión electrónica http://www.inabima.org/BibliotecaInabima2/ASIGNATURAS/F%CDSICA/Introduccion%20a%20la%20fisica%20cuantica.pdf. Consultado por última vez 06/09/2010].

RIOJA NIETO, Ana María. "Los orígenes del principio de indeterminación". *Theoria: Revista de teoría, historia y fundamentos de la ciencia*, 10 (22), 1995, pp. 117-142.

RIVADULLA, Andrés. *Éxito, razón y cambio en física: un enfoque instrumental en teoría de la ciencia.* Madrid: Trotta, 2004.

RIVADULLA, Andrés. *Revoluciones en física.* Madrid: Trotta, 2003.

ROSENBERG, Alex. "Reductionism in a Historical Science". *Philosophy of Science,* 68 (2), 2001, pp. 135-163.

ROSENBERG, Alex. "Reductionism redux: computing the embryo". *Biology and Philosophy,* 12 (4), 1997, pp. 445-470.

SALAS PERALTA, Pedro J. y SANZ SÁENZ, Ángel L. "Corrección de errores en ordenadores cuánticos". *Revista Española de Física (REF),* 20 (1), 2006, pp. 20-27.

SARKAR, Sahotra y PLUTYNSKI, Anya (Eds.). *A Companion to the Phylosophy of Biology.* Oxford: Blackwell Publishing, 2007.

SARKAR, Sahotra. *Genetics and Reductionism.* Cambridge: University Press, 1998.

SARKAR, Sahotra (Ed.). *The philosophy and history of molecular biology: new perspectives.* Dordrecht, London: Kluwer Academic, 1996.

SCHRÖDINGER, Erwin. *¿Qué es la vida?.* Traducción y notas de Ricardo Guerrero. Barcelona: Tusquets, 2008.

SCHUMACHER, Benjamin. "Quantum coding". *Physics Review A.*, 51, 1995, pp. 2738-2747.

SHANNON, Claude. "Communication theory of secrecy systems". *Bell Systems Technical Journal*, 28, 1949, pp. 656-715.

SHANNON, Claude. *"*A mathematical theory of communication". *Bell Systems Technical Journal*, 27, 1948, pp. 379-423, 623-565.

SHAPIRO, Robert. "El origen de la vida". *Investigación y ciencia,* 371, 2007, pp. 19-25.

SHAPIRO, Robert. "Small molecule interactions were central to the origin of life". *The Quarterly Review of Biology,* 81 (2), 2006, pp. 105-125.

SKLAR, Lawrence. *Filosofía de la física.* Versión española de Rosa Álvarez Ulloa. Madrid: Alianza, 1994.

SMITH, Caroline, Russell, Sara y Benedix, Gretchen. *Meteorites*. London: Natural History Museum, 2009.

STANFORD Encyclopedia of Philosophy http://plato.stanford.edu [Consultado por última vez 07/09/2010]

STEANE, Andrew Martin. "Error correcting codes in quantum theory". *Physical Review Letters*, 77 (5), 1996, pp. 793-797.

SUGAWARA, Hirotaka. "Quantum theory of the DNA". [http://ptp.ipap.jp/link?ptps/164/17. Consultado por última vez el 10/09/2010]

TORRETTI, Roberto. *The Philosophy of Physics.* Cambridge: University Press, 1999.

WÄCHTERSHÄUSER, Günter. "The Origin of Life and its Methodological Challenge". *Journal of Theoretical Biology,* 187 (4), 1997, pp. 483-494.

WORKSHOP OQOL'09: Open Questions on the Origins of Life: San Sebastián-Donostia, May 20-23, 2009: Book of Abstracts (First Draft). Pier Luigi Luisi & Kepa Ruiz-Mirazo (org.)

# GLOSARIO

- **Atmósfera reductora:** es aquella compuesta por gases que ceden electrones.
- **Autopoyesis:** continua producción y automantenimiento de los sistemas vivos.
- **Bit:** es un dígito del sistema de numeración binario
- **Catalizador:** compuesto que acelera las reacciones químicas
- **Clonación de estados:** replicaciones exactas de estados.
- **Colapso del estado:** proyección del estado cuando se realiza la medida
- **Cuanto:** valor mínimo que puede tomar una magnitud, i.e. energía
- **Decoherencia:** pérdida de la propiedad de superposición y entrelazamiento
- **DNA:** tipo de ácido nucleico con estructura de doble hélice
- **Electrón:** partícula subatómica con carga negativa.
- **Espacio vectorial:** es un conjunto, i.e vectores, que cumple ciertas relaciones de suma y producto.
- **Espín:** en las partículas subatómicas, capacidad de giro o momento angular.
- **Estado entrelazado/enredado:** estado en el que se encuentran interrelacionadas dos o más partículas.
- **Entropía:** medida del desorden del sistema.
- **Enzima:** biocatalizador.
- **Fotón:** cuanto de luz
- **Homeostasis:** situación estable o de equilibrio.
- **Ley del movimiento:** relación funcional entre la posición, la cantidad de movimiento y el tiempo
- **Nucleótido:** moléculas orgánicas formadas por la unión covalente de un monosacárido (pentosa), una base nitrogenada y un grupo fosfato.
- **Observable:** toda propiedad del estado de un sistema que puede ser determinada por alguna secuencia de operaciones físicas.
- **Preparar un sistema**: situar el sistema en unas condiciones iniciales experimentales
- **Qubit:** es un sistema cuántico con dos estados propios que puede tener un continuo de estados. Análogo cuántico al bit.
- **RNA:** tipo de ácido nucleico
- **Reducción de un estado:** medida
- **Superposición de estados:** un objeto "posee simultáneamente" varios valores de una cantidad observable.
- **Tautómero:** dos moléculas que se diferencian sólo en la posición de un grupo funcional.

**APÉNDICE I**

**Postulados de la Mecánica Cuántica:**

**Postulado I**

En un instante de tiempo *t,* un sistema cuántico está totalmente determinado por un vector del espacio de estados: $|\psi> \in$ al espacio de estados $\xi$

- Principio de superposición: si tenemos varios estados posibles, una combinación lineal de estados también es estado del sistema

Consecuencia de que el espacio de estados es lineal, entonces se cumple el principio de superposición: $|\psi_1> y |\psi_2> \in$ al espacio de estados $\xi \rightarrow \alpha|\psi_1> + \beta|\psi_2> \in \xi$

La norma del estado debe ser 1, $|||\psi>|| = 1$, que es la probabilidad unidad. Pero no hay definición unívoca del estado pues la norma sería unidad salvo en una fase $|\psi> \rightarrow e^{\theta i}|\psi>$ entonces $|||\psi>|| = 1$ y $||e^{\theta i}|\psi>|| = 1$

**Postulado II**

A toda magnitud física $O$ le hacemos corresponder un observable $A$ (operador) que actúa sobre el espacio de estados $\xi$ con $A$ hermítico. De esta forma, podemos construir una base de estados $O \rightarrow A$ (*observable*)

**Postulado III**

Si se tiene una magnitud física $O$ descrita por $A$; cuando se mide la magnitud física, los únicos valores posibles son los autovalores del observable $A$ asociados a la magnitud física.

- *Como A es hermítico $\rightarrow \lambda$ son autovalores reales*
- *Si el espectro de autovalores es discreto $\rightarrow$ las medidas están cuantizadas*
- *Si $|\psi>$ es el estado del sistema, no podemos en general saber el valor que vamos a obtener, solo sabemos que es uno de los autovalores de A.*
- Si un sistema está en el estado $|\psi>$ donde $<\psi|\psi> = 1$ con un observable $A|u_n> = a_n|u_n>$ con $A$ conjunto completo de observables que conmutan, $a_n$ discreto $\rightarrow$ *los $|u_n>$ son base discreta* y $|\psi> = \sum c_n |u_n> = \sum <u_n|\psi>|u_n>$

**Postulado IV**

Si al medir el estado $|\psi>$ obtengo $a_n$ cualquier autovalor es posible pero no todos se obtienen con la misma probabilidad.

- $P(a_n) = $ *probabilidad de obtener $a_n$ al hacer la medida* $= |c_n|^2 = |<u_n|\psi>|^2$

    *con $|u_n>$ autovector asociado al autovalor $a_n$*

*La probabilidad de encontrar un cierto autovalor es la componente de la proyección de $|\psi>$ sobre el autovect $|u_n> \rightarrow$ si calculamos $\sum P(a_n) = \sum |<u_n|\psi>|^2 = \sum <u_n|\psi><u_n|\psi>^*$*

$$= \sum <u_n|\psi><\psi|u_n> \quad = \sum |u_n><u_n||\psi><\psi| = 1$$

Si $A$ no es un conjunto completo de observables que conmutan, entonces hay degeneración del sistema $A|u_n^i> = a_n|u_n^i>$ y $\xi_n$ es el subespacio lineal formado por los autovectores de $A$ con autovalores $a_n$ con degeneración $i = 1, 2, \dots g_n$ por lo que $|\psi> = \sum_{i=1,2}^{g_n} c_n^i |u_n^i>$

$$\sum P(a_n) = \sum_{i=1}^{g_n} |<u_n^i|\psi>|^2 = \sum_{i=1}^{g_n} |c_n^i|^2 = 1$$

El proceso de medida sigue la siguiente secuencia tempora

*$|\psi_0> \rightarrow$ evolución temporal según la ecuación de Schrödinger $\rightarrow |\psi_t>$*

*se produce la medida de A con autovalores $a_n$ en $t_m \rightarrow$*
*en este momento se ha proyectado sobre$|u_n> \rightarrow$*
*y el sistema sigue evolucionando siguiendo la ecuación de*
*Schrödinguer*

De esta forma vemos que el proceso de medida cambia el estado $|\psi> \rightarrow P_{proyector}|\psi> = |u_n>$

**Postulado V**

Si la medida de la magnitud asociada al observable $A$ es $a_n$ entonces, el estado después de la medida pasa a ser $\frac{P_{proyector}|\psi>}{N}$ donde $N$ es el factor de normalización. Es decir $|\psi> \rightarrow \frac{P_{proyector}|\psi>}{<\psi|\psi>}|$

**Postulado VI**

Mientras no se realiza ninguna medida, la evolución en el tiempo del vector de estado del sistema viene determinada por la ecuación de Schrödinger:

$i\frac{h}{2\pi}\frac{d}{dt}|\psi> = H(t)|\psi>$ donde $H(t)$ es el observable asociado a la energía total del sistema. Si el sistema está aislado, la evolución es determinista siguiendo la ley ecuación de Schrödinger. El indeterminismo aparece al realizar la medida puesto que no sabemos el autovalor, ni por lo tanto el autovector o subespacio del sistema.

Las condiciones iniciales se pueden preparar midiendo un conjunto completo de observables que conmuten.

**Reglas de cuantización:**

i. *conmutan* $[R,P] = 0$

ii. *no conmutan* $[R,P] = i\frac{h}{2\pi}\delta$

Si una magnitud $O$ viene definido por una relación $f(r,p,t)$ enonces le podemos asignar un observable y operador $A(t) = f(R,P,T)$ que se puede obtener, por ejemplo, mediante desarrollo de Taylor. Se debe hacer notar que no siempre los observables cuánticos vienen de magnitudes clásicas asociadas, por ejemplo el espín.

**APENDICE II**

**Efecto túnel:**

En la física clásica se sabe que para que un patinador logre superar la cima de una cuesta debe llevar una velocidad suficiente para que su energía cinética asociada le permita alcanzar la altura de la de la cuesta. Es decir, de acuerdo con la física clásica, si la energía de una partícula es inferior que la altura de cierta barrera de potencial, no hay forma de traspasar dicha barrera de potencial (fig. 1).

En el mundo cuántico, el estado de la partícula, pongamos un protón, viene descrito por su función de onda y ésta debe ser continua tanto a la entrada como a la salida de la barrera de potencial. Esto significa que mostrará la posibilidad de traspasar la barrera: éste es el denominado efecto túnel (fig.2).

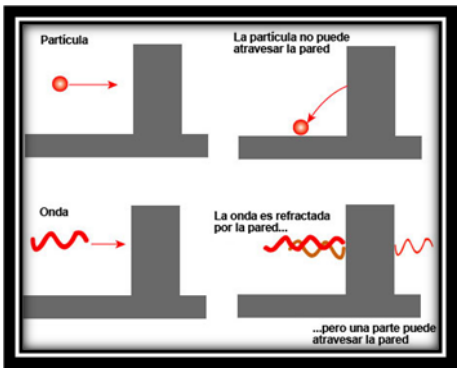
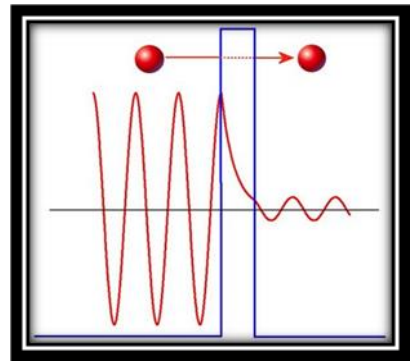

Fig. 1                      Fig. 2

En el caso del efecto túnel protónico, los autores proponen un modelo cuántico unidimensional de dos núcleos representados por dos barreras de potencial en forma de pozo. Uno de los protones se encuentra en superposición a los dos lados de los pozos mencionados antes de la medida. El potencial de oscilación, que es la energía que mantiene ligado al protón entorno a los dos núcleos, depende de la distancia entre las dos localizaciones del núcleo $2a$ con $a \sim 1A$ de la masa del protón y de la frecuencia de oscilación de éste ($\omega = 10^{12} Hz$). Si se obtienen los valores de la energía más bajos, se puede utilizar la fórmula $\Delta E \Delta t \geq h/2\pi$ para calcular $\Delta t \geq h/2\pi \Delta E$ que nos proporciona una estimación del tiempo que necesita una partícula para ir de un lado del pozo de potencial al otro ($\approx 10^{12} s$).

# APENDICE III

## Efecto Zenón

Zenón de Elea solía defender las tesis de su maestro Parménides mediante paradojas sobre las contradicciones que presentaba el movimiento y la multiplicidad del ser. En el caso que nos ocupa, si se piensa en una flecha en movimiento que ocupa a cada instante una posición bien definida esto nos lleva a la contradicción de que no puede estar en movimiento.

La trayectoria bien definida significa un camino suave, que en matemáticas se denomina trayectoria continua. En principio, en este tipo de trayectorias se puede calcular la velocidad de una partícula, que es la relación entre el espacio recorrido en un intervalo de tiempo $v = \frac{\Delta s}{\Delta t}$. Si se desea alcanzar más precisión, se debe tomar el intervalo de tiempo cada vez más pequeño. Si se toma ese incremento de tiempo suficientemente pequeño, se constata que la velocidad deja de tender a un límite $v \neq \lim_{\Delta t \to 0} \frac{\Delta s}{\Delta t} = \frac{ds}{dt}$ ya que lo impiden los choques de las moléculas (movimiento browniano) con nuestra partícula. Si se isla bien el sistema para que no sufra interacciones con las moléculas del entorno, entonces el límite tiende a la velociadad de la partícula $v = \lim_{\Delta t \to 0} \frac{\Delta s}{\Delta t} = \frac{ds}{dt}$.

Ahora bien, en mecánica cuántica también falla la descripción continua de la trayectoria. Es decir, el concepto de trayectoria continua no se aplica al movimiento de las partículas. Realmente, si se recuerda el principio de indeterminación de Heisenberg, no se puede definir la trayectoria de la partícula sin perder toda la información sobre su momento.

Como se vio en el apéndice I, la evolución de un sistema, mientras no experimenta la medida, viene dada por la ecuación de Schrödinger $i\frac{h}{2\pi}\frac{d}{dt}|\psi> = H(t)|\psi>$.

Si se tiene un sistema que decae con el tiempo, por ejemplo las transiciones de un átomo excitado, y le sometemos a la medida, el sistema colapsa $|\psi> \to P_{proyector}|\psi> = |u_n>$. Si hacemos un gran número de medidas en tiempos muy cortos, el estado quedará "congelado" en el estado inicial y no se producirá la transición entre estados. Por lo tanto, como la decoherencia del estado no es instantánea, si tenemos un sistema en contacto con el entorno se producirá la medida en un periodo corto de tiempo debido al fuerte acoplamiento entre el sistema y el entorno. Si el acoplamiento con el entorno es continuo, esto equivale a realizar frecuentes medidas. Cuanto mayor es el acoplamiento, y más pequeño es el tiempo de decoherencia, más rápido se producirá el colapso. Así, si el sistema está fuertemente acoplado con el entorno y este acoplamiento se produce de forma continua, la evolución del sistema se detiene. El efecto Zenón se ha medido experimentalmente en varias situaciones (2001, Mark G. Raizen, K. Koshino y A Shimizu (2005))

Con más precisión: si $A$ es un observable donde $A|u_n> = a_n|u_n>$ con $|u_n>$ autoestados y $a_n$ autovalores asociados. Si obtenemos el estado del sistema $|\psi>$ al realizar la medida con auto valor $a_n$:

$$P(a_n) = probabilidadde\ obtener\ a_n\ al\ hacer\ la\ medida = |<u_n|\psi>|^2$$

cuando obtenemos ese valor $a_n$ el estado del sistema después de la medida es $|u_n>$.

Si el sistema en el instante inicial se encuentra en el autoestado $|u_n>$ del observable $A$ : $|\psi>= e^{Ht/i\hbar}|u_n>$. Donde $H$ es la hamiltoniana del sistema. Entonces la probabilidad de obtener el valor $a_n$ es $P(a_n) = |<u_n|\psi>|^2$.

Si el intervalo de tiempo es muy pequeño, entonces la evolución temporal del sistema $e^{Ht/2\pi i\hbar}$ puede desarrollarse en serie $e^{Ht/2\pi i\hbar} = \left(1-\frac{H^2\Delta t^2}{2\hbar^2}\right) + i\hbar\Delta tH$ sustituyendo $|\psi>= \left(1-\frac{H^2\Delta t^2}{2\hbar^2}\right)|u_n> + i\hbar\Delta tH |u_n>$. La probabilidad de obtener $a_n$ será $P(a_n) = \left(1-\frac{\Delta_n^2 H\Delta t^2}{\hbar^2}\right)$ donde

$\Delta_n^2 H = <u_n|H^2|u_n> - <u_n|H|u_n>^2$. Si medimos un número $N$ de veces obteniendo $a_n$ entonces $P(N, a_n) = \left(1-\frac{\Delta_n^2 H\Delta t^2}{\hbar^2}\right)^N$ donde el sistema evoluciona entre mediciones según la ecuación de Schrödinger y siempre colapsa al estado inicial $|u_n>$. Si $N \to \infty \ e\ \Delta t \to 0; N\Delta t \to t$ lo que implica que $\lim_{N\to\infty\ e\ \Delta t\to 0} P(N, a_n) = 1$. Por lo tanto: el sistema permanece congelado sin experimentar transiciones ni decaer.

**APENDICE IV**

**Efecto Casimir**

Este efecto está relacionado con el concepto de vacío de la teoría cuántica de campos. Un campo clásico puede representarse mediante un conjunto de partículas elementales engastadas en muelles de forma continua. La excitación del sistema pone en movimiento las partículas con sus muelles. En un sistema cuántico la energía de los muelles no se transmite de forma continua sino discreta. Si representamos el vacío cuántico como aquella región donde las propiedades (espín, polarización, carga,…) de las partículas se cancelan en promedio. Al alterar este equilibrio del vacío, situando, por ejemplo, en torno a éste dos conductores a una distancia pequeña comparada con su magnitud, entonces cambiará la energía del campo magnético cuantizado, de tal suerte que aparecerá una fuerza atractiva entre ellos debido a la alteración del vacío cuántico. Esta fuerza es importante a escalas de $1\mu m$ y domina a escalas de $10^{-8} m = 10 nm$.

El efecto Casimir es muy importante en la quirilidad del núcleo, es decir en la elección de dos configuraciones simétricas equivalentes, y se aprecia en las fuerzas de enlace de Van der Waals entre un par de átomos neutros. Steve K. Lamoreaux y Umar Mohideen midieron este efecto en 1997. Un grupo de la Universidad de Padua volvió a corroborar este efecto en el año 2001.

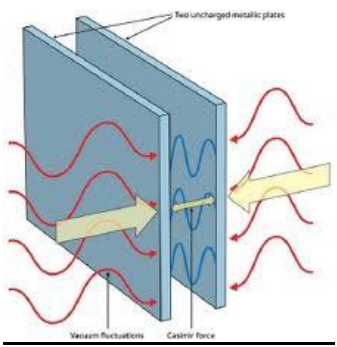

**APÉNDICE V**

**Moléculas en el espacio**

Desde mediados de los años treinta del siglo pasado, se conoce la existencia de algunas moléculas en el espacio (CN, CH, CH$^+$), tanto en el medio interestelar como formando parte de los distintos objetos celestes. Sin embargo, fue a través de las nuevas técnicas radio astronómicas en los años sesenta cuando la astrofísica molecular produjo un cambio sustancial, cualitativo y cuantitativo, en la manera de entender la formación de moléculas en el espacio. Sirva como ejemplo el descubrimiento de vapor de agua en 1969.

La formación de moléculas en entornos extremos de densidad y temperatura no tuvo una explicación convincente hasta épocas muy recientes, debido a que la química que se utilizaba para describir dicha formación no era la debida, por ejemplo: en la Tierra la formación de moléculas se consigue a través de interacciones a tres cuerpos, mientras que en el espacio se realiza mediante interacciones a dos cuerpos.

Las moléculas en el espacio se observan en todas las direcciones con tal de que la temperatura sea inferior a $2000 K$ y en entornos con densidades superiores a $100\, part/cm3$. Se han encontrado unos $130$ tipos de moléculas que coinciden en gran parte con las que ya conocíamos en la Tierra. También se han detectado moléculas no observadas en laboratorios terrestres ($C_{60}$) debido a su alta inestabilidad. El espacio muestra una vez más su potencialidad como laboratorio en condiciones extremas.

Los nuevos proyectos como el ALMA ( una red interferométrica de varios telescopios) que tendrá una resolución como en el óptico para fuentes en el infrarrojo, nos ayudarán a profundizar en este campo que todavía promete muchas sorpresas.

Se han encontrado moléculas en cuásares con alto desplazamiento al rojo     ( $CO$ en QSO a Z=4.7), en galaxias de todos los tipos, incluso en las elípticas.

Parte de la masa de la galaxias está en forma de estrellas y el resto, que es una cantidad muy significativa, está en forma de gas tanto en hidrógeno atómico $H$ como en $H_2$ molecular.

El medio interestelar (ISM) tiene densidades del orden de 1.23 cm$^{-3}$ y temperaturas cinéticas bajas. No obstante, hay regiones de densidad extrema, por ejemplo hay nubes que pueden exceder las $1000 Mo$ y regiones con temperaturas cinéticas del orden de $10^5 - 10^6\, K$ debido a la existencia de fuentes ionizantes, como estrellas tipo A,O..., a ondas de choque debidas a estrellas en fase de supernova, y a la baja densidad del entorno.

En los entornos de formación estelar, como las regiones HII (cuyo ejemplo más conocido es la nebulosa de Orión), las moléculas son visibles en infrarrojo (vapor de agua) y microondas. Algunas emiten incluso en radio ondas (máser), mientras que otras son fuentes enormes de radiación térmica a bajas frecuencias.

Los fotones de las fuentes de alta ionización disocian las moléculas de la cáscara superficial guardando así en su interior moléculas que son observadas a través de líneas rotacionales que sirven de trazadores del gas frío.

Las nubes oscuras del ISM son fuente frecuente de emisión debido a transiciones rotacionales de grandes moléculas orgánicas como lo hidrocarburos policíclicos aromáticos (PAH,s).

En las regiones circumestelares de estrellas evolucionadas y en regiones HII compactas se forman los granos polvo que juegan un papel muy importante en la creación y estabilidad de moléculas en el ISM. Las abundancias moleculares de estas regiones vienen condicionadas por la riqueza en C u O de las estrellas de su entorno.

Es evidente la existencia y abundancia de especies moleculares en lunas y planetas, buen ejemplo de ello es la vida en la Tierra, las lluvias de acetileno en Titán y las imágenes impresionantes del metano eyectado en el choque del cometa Sumaker-Levi contra Júpiter. Las observaciones de líneas moleculares nos dan información sobre la estructura de las atmósferas planetaria y de parámetros físicos como la temperatura.

Los cometas nos aportan datos sobre la nube parental del Sistema Solar, a través de las moléculas formadas con átomos de composición isotópica primitiva. Esto se debe a que en las moléculas con isótopos diferentes, por ejemplo $Li^6$ y $Li^7$, cambia el espín nuclear afectando a la estructura fina del espectro molecular. Su composición molecular es de alta relevancia para el problema del surgimiento de la vida:

| Componentes del meteorito Murchinson | Abundancias µg g 1 (ppm) |
|---|---|
| Material Macromolecular | 1.45 (%) |
| Dióxido de Carbono | 106 |
| Monóxido de Carbono | 0,06 |
| Metano | 0,14 |
| Hidrocarburos: alinfáticos | 12 – 35 |
| aromaticos | 15 – 28 |
| Ácidos carboxilico | 332 |
| dicarboxilico | 25,7 |
| α-hidrocarboxilico | 14,6 |
| Aminoácidos | 60 |
| Alcoholes | 11 |
| Aldeidos | 11 |
| Cetonas | 16 |
| Componentes rel.azucar | 60 |
| Amonio | 19 |
| Aminas | 8 |
| Urea | 25 |
| (piridinas, quinolinas) | 0,05-0,5 |
| Pyridinecarboxylic acids | 7 |
| Dicarboximides | 50 |
| Uracilo y timina | 0,06 |
| Purinas | 1,2 |
| Benzotiofenos | 0,3 |
| Ácido sulfónico | 67 |
| Ácido fosfónico | 1,5 |

Las estrellas en sí mismas son ricas en moléculas (por ejemplo, en el Sol se observan las bandas del vapor de agua con claridad).

En el siguiente cuadro se muestran algunas características de los distintos objetos del ISM.

| Objeto | Temp. K | Den. cm$^{-3}$ | Atomos y moléculas |
|---|---|---|---|
| Región HII | 10000 | 100-1000 | Iones de $H, C, N, O$ |
| Nube difusa | 100 | 100 | Ion C, Moléculas $CO, H2CO$... |
| Nube oscura | 10-20 | 10000 | Muchos tipos moleculares: $CH_3CH_4$ ; $CH_2OHCH_2$ |
| Nube molecular | 50 | I00000 | Muchos tipos moleculares: $CH_2CHCN$; $HC_{11}N$ |
| Región circunestelar | 100-1000 | 100-1000 | Polvo, en estrellas oxigenadas gran abundancia de $SiO$, en carbonadas $CO, C2\,H2$ |
| Reg. HII compacta | 100-1000 | 1000-10000 | Máseres de vapor de agua $SiO, OH$ |

# APÉNDICE VI

## La ley de Stephan Boltzman-Shanon

El físico suizo relacionó la probabilidad con la termodinámica del sistema asumiendo que las probabilidades de distribución de las microconfiguraciones del sistema correspondían a estados macroscópicos del sistema. Así, el número de maneras de conseguir una configuración dada está relacionado con la probabilidad de obtener dicha distribución. Si llamamos probabilidad termodinámica $W$ al número de maneras de obtener una configuración, entonces este número estará comprendido entre cero e infinito. Pero la probabilidad siempre se encuentra acotada entre 0, o imposibilidad, y uno, o certeza. Por lo tanto si $S$ representa el grado de desorden y $W$ es la probabilidad, entonces existe una función que relaciona $S$ con $W$. Si la entropía se comporta como la masa o el volumen, es decir, si dos sistemas con diferentes entropías se juntan y la entropía final del sistema es la suma de las dos, es decir si el grado de desorden del conjunto es la suma de los desórdenes de los dos sistemas

[9] $S = f(W) = f(W_1) + f(W_2)$

entonces la función que relaciona el grado de desorden con la probabilidad de los dos sistemas es la suma de las funciones respectivas de cada uno de los sistemas:

[10] $S = f(W) = f(W_1) + f(W_2)$

Pero si las probabilidades de cada sistema influyen la una en la otra entonces la probabilidad total será el producto de las probabilidades $W = W_1 \times W_2$ y por lo tanto la función de la probabilidad total será función del producto:

[11] $S = f(W) = f(W_1 \times W_2)$

La función que cumple [10] y [11] a la vez es el logaritmo que además que para valores de $W$ entre 1 e infinito toma valores entre 0 y 1 es decir tiene forma de probabilidad.

[12] $S \propto \ln W$

De igual manera, la información $I$ es función de la probabilidad descrita por la razón de los casos favorables a los posibles. $I = f(P)$ como las probabilidades cumplen la regla del producto, entonces siguiendo el análisis descrito anteriormente:

[13] $I \propto \ln P$ en unidades de información o bits